# 9 Saturn's Variable Thermosphere

Darrell F. Strobel

*Departments of Earth & Planetary Sciences and Physics & Astronomy, Johns Hopkins University, Baltimore, MD 21218, U.S.A.*

Tommi Koskinen

*Lunar and Planetary Laboratory, University of Arizona, Tucson, AZ 85721-0092*

Ingo Müller-Wodarg

*Blackett Laboratory, Imperial College, London, UK*

**ABSTRACT**

Our knowledge of Saturn's neutral thermosphere is far superior to that of the other giant planets due to Cassini Ultraviolet Imaging Spectrograph (UVIS) observations of 15 solar occultations and 26 stellar occultations analyzed to date. These measurements yield $H_2$ as the dominant species with an upper limit on the H mole fraction of 5%. Inferred temperatures near the lower boundary are ~ 150 K, rising to an asymptotic value of ~ 400 K at equatorial latitudes and increasing with latitude to polar values in the range of 550-600 K. The latter is consistent with a total estimated auroral power input of ~ 10 TW generating Joule and energetic particle heating of ~ 5-6 TW that is more than an order of magnitude greater than solar EUV/FUV heating. This auroral heating would be sufficient to solve the "energy crisis" of Saturn's thermospheric heating, if it can be efficiently redistributed to low latitudes. The inferred structure of the thermosphere yields poleward directed pressure gradients on equipotential surfaces consistent with auroral heating and poleward increasing temperatures. A gradient wind balance aloft with these pressure gradients implies westward, retrograde winds ~ 500 m s$^{-1}$ or Mach number ~ 0.3 at mid-latitudes. The occultations reveal an expansion of the thermosphere peaking at or slightly after equinox, anti-correlated with solar activity, and apparently driven by lower thermospheric heating of unknown cause. The He mole fraction remains unconstrained as no Cassini UVIS He 58.4 nm airglow measurements have been published.

## 9.1. INTRODUCTION

Traditionally the thermosphere is defined as the region characterized by a steep temperature gradient generated at its base by intense heating from absorption of short wavelength solar ultraviolet radiation (< 170 nm) in the dissociation and/or ionization of homonuclear molecules, e. g. $H_2$, $N_2$) and the downward transport of thermal energy by heat conduction due to the absence of infrared active molecules. The base of the thermosphere is the mesopause defined as the level where the temperature reaches a minimum and the temperature gradient vanishes due to infrared active molecules there radiating away the solar UV heating above. As with all solar system giant planets, however, solar UV heating is a minor energy source for Saturn's thermosphere.

The thermosphere is typically located above the homopause, where atmospheric species undergo gravitational diffusive separation and assume scale heights in accordance with their atomic and molecular masses rather than the atmospheric mean. Methane density profiles as determined from occultations provide relatively precise locations of the $CH_4$ homopause in contrast to the much larger uncertainty in using inferred temperature profiles to locate the base of thermosphere.

An authoritative chapter on Saturn's thermosphere is challenging for one cannot replicate the continuous vertical structure achieved by the Galileo probe at Jupiter or the Huygens probe at Titan. Over some altitude regions, temperature, composition, and density profiles are retrieved as a function of pressure, whereas for others data are acquired as a function of radial distance or, in particular near the mesopause, are only marginally available. A definitive He/H2 mixing ratio, which would enable construction of continuous atmospheric profiles at individual locations, is also lacking at Saturn, for which we have the value of 0.034 inferred from Voyager 1 infrared and radio occultation measurements, universally regarded as incorrect and at best an extreme lower bound. Instead, Conrath and Gautier (2000) inferred a ratio of ~ 0.13 from reanalysis of Voyager IRIS data.

The chapter is organized to establish what we know or at least what we think we know from density data (Section 2), temperature data (Section 3), airglow data (Section 4), and composition data (Section 5), followed by an examination of the density, pressure, and temperature structure derived from the available data and the inference of net heating rates from radial temperature profiles (Section 6). In Section 7, UVIS stellar occultation data are used to infer the location of the $CH_4$ homopause. Energetics and potential sources are discussed and reviewed in Section 8, which sets the stage for Section 9 on global general circulations models for the thermosphere and a discussion of the observed state of Saturn's atmosphere as compared to the models. Auroral physics and ionospheric chemistry and physics, often considered part of thermospheric physics, are dealt with in Chapters 7 and 8. We treat them here only as necessary for our purposes.

## 9.2. REVIEW OF RELEVANT DENSITY DATA

The neutral thermosphere is composed mainly of $H_2$ and He with trace amounts of H and water group molecules (O, OH, $H_2O$). Atomic H is released by photochemical reactions below the $CH_4$ homopause (e.g., Moses et al. 2005) and its mixing ratio increases with altitude in the thermosphere due to molecular diffusion. Water group molecules are delivered to the upper atmosphere from Saturn's magnetosphere and rings (e.g., Connerney and Waite 1984; Feuchtgruber et al. 1997; Cassidy et al. 2010; O'Donoghue et al. 2013). Their net mixing ratio is expected to decrease with increasing pressure in the stratosphere due to condensation in the lower atmosphere and be roughly constant with altitude in the upper atmosphere (e.g., Moses et al. 2000; Müller-Wodarg et al. 2012). The mixing ratios of $CH_4$ and other hydrocarbons, on the other hand, decrease rapidly with altitude above the homopause due to molecular diffusion and photolysis, and should be negligible in the thermosphere. The mixing ratio of He also decreases with altitude in the thermosphere, but it does so less rapidly than the mixing ratios of the hydrocarbons.

Observational constraints on $H_2$ and H can be obtained from extreme (EUV) and far-ultraviolet (FUV) occultations of bright stars and the Sun as well as airglow and auroral emissions (e.g., Broadfoot et al. 1981; Sandel et al. 1982; Smith et al. 1983; Shemansky and Ajello 1983; Yelle 1988; Gérard et al. 1995, 2009, 2013; Shemansky et al. 2009; Gustin et al. 2010). Constraints on He can be obtained from EUV airglow measurements (Sandel et al. 1982; Parkinson et al. 1998) while $CH_4$ near the homopause can be probed by FUV occultations (Smith et al. 1983; Shemansky and Liu 2012; Koskinen et al. 2015; Vervack and Moses 2015) as well as limb emissions and occultations in the near-IR $CH_4$ bands at 3.3 μm (Baines et al. 2005; Kim et al. 2012).

Unfortunately there are no direct observations of the water group molecules reported at present even though they may play a substantial role in controlling the electron densities in the ionosphere (e.g., Moore et al. 2010; Müller-Wodarg et al. 2012, cf. Chapter 8).

In principle, stellar and solar occultations are the most reliable source of information about the density and temperature structure in the thermosphere.  This is because the interpretation of airglow and auroral emissions depends on free parameters in complex non-LTE radiative transfer models (e.g., Liu and Dalgarno 1996; Hallett et al. 2005; Gustin et al. 2010; García-Comas et al. 2011; Adriani et al. 2011) and instrument calibration that is not always well characterized.  The analysis of the occultations, on the other hand, is much simpler.  The data do not need to be absolutely calibrated and the retrieved densities are simply based on the line of sight optical depths of the absorbers that can be obtained from the raw data with relative ease.  The occultations also probe the atmosphere at a wide range of altitudes and latitudes.  In this context, it may appear surprising that there are significant disagreements between different analyses of the Voyager/UVS (Broadfoot et al. 1981; Festou and Atreya 1982; Vervack and Moses 2015) and Cassini/UVIS occultations (Shemansky and Liu 2012; Koskinen et al. 2013, 2015).  The recent re-analysis of the Voyager/UVS data (Vervack and Moses 2015) and new results based on 41 solar and stellar occultations from Cassini/UVIS (Koskinen et al. 2013, 2015) have resolved this dispute.

### 9.2.1 Voyager/UVS occultation data

Voyager 1 (V1) observed two solar occultations and one stellar occultation of ι Herculis while Voyager 2 (V2) observed one solar occultation and two stellar occultations of δ Scorpii.  Density and temperature profiles retrieved from all six of the Voyager/UVS occultations were published only recently by Vervack and Moses (2015).  Previous results were limited to exospheric temperatures and simple forward models of the atmosphere based on only three of the occultations (Broadfoot et al. 1981; Sandel et al. 1982, Festou and Atreya 1982; Smith et al. 1983; Shemansky and Liu 2012).  There are significant differences in the retrieved exospheric temperatures between these results that we discuss further in Section 9.3.2.  Here we concentrate only on the newly published density profiles of $H_2$ and H (Vervack and Moses 2015).

In order to better understand the results and their limitations, it is useful to develop a basic understanding of the instruments and the occultation data. The processed data consist of transmission spectra as a function of radial distance and latitude from the center of Saturn. Together these spectra make up light curves i.e., transmission as a function of radial distance in each wavelength band. The wavelength range of the Voyager/UVS instruments is 51 – 170 nm with a spectral resolution of 1.8 – 3 nm for point sources. The signal-to-noise (S/N) of the occultation data, however, was often so poor that in reality the data had to be binned to a much lower resolution. The observations also suffered from significant spacecraft pointing drifts and slews, interference from the rings, changes in detector gain and pixel-to-pixel sensitivity variations. All these effects significantly complicated the retrieval of density profiles from the Voyager/UVS occultations (Vervack and Moses 2015).

The density profiles of $H_2$ and H are retrieved by analyzing transmission at wavelengths of 51 – 115 nm. Hydrogen in the interstellar medium (ISM) absorbs all starlight at wavelengths shorter than 91 nm and thus the shorter wavelengths in this range are only available in the solar occultations. This is an important point because absorption at 51 – 80.4 nm is dominated by the ionization continuum of $H_2$. The absorption cross section in the ionization continuum varies smoothly with wavelength and, even more importantly, does not depend on temperature (Samson and Haddad 1994). This is in contrast to the electronic Lyman and Werner bands of $H_2$ (hereafter, the LW bands) at longer wavelengths where the cross section is much more complicated and depends on temperature. Thus solar occultations provide the most robust measurements of the density profiles and exospheric temperatures. The results, however, are limited to relatively high altitudes near the exobase above the 0.1 nbar level. An analysis of the LW bands is still required to probe the density and temperature profiles in the lower thermosphere and near the homopause (~ 0.01-0.1 µbar). The exobase is formally defined as the level in the atmosphere where the mean free path of the major species ($H_2$) is equal to its scale height divided by $\sqrt{2}$ and regarded as the top of the atmosphere (Strobel, 2002). If the collisionless, neutral gas above the exobase is gravitationally bound, it is referred to as the exosphere.

Figure 9.1 shows the $H_2$ and H profiles based on the V1 and V2 data from Vervack and Moses (2015), together with results b from the previous analysis of the V2 occultations by Smith et al. (1983) and ground-based observations of the occultation of the star 28 Sgr (Hubbard et al. 1997). Vervack and Moses (2015) retrieved the lowest pressure $H_2$ densities (red line) from the ionization continuum in the V2 solar occultation while the rest of the $H_2$ profiles below the 0.1 nbar level are based on the LW bands. They used the same canonical T-P profile to calculate the absorption cross sections for all occultations and did not account for the change in temperature along the line of sight in this calculation. Given the relatively large uncertainties in the Voyager/UVS light curves, however, these assumptions are unlikely to result in significant systematic errors in the retrieved density profiles. As illustrated by Figure 9.1, the retrieval of the $H_2$ profiles is typically limited to altitudes above the 10 nbar level – a level where transmission in the LW bands vanishes. Owing to the sophisticated treatment of the instrument effects and updated retrieval techniques used by Vervack and Moses (2015), the density profiles based on the Voyager/UVS data are now remarkably consistent. They also agree well with the density profile retrieved by Hubbard et al. (1997) at all relevant altitudes.

The abundance of H in Figure 9.1 is retrieved from the Lyman continuum at 80.4 – 91 nm in the V1 and V2 solar occultations. The analysis of the higher resolution solar occultations from Cassini/UVIS (see Section 2.2 below) suggests that separating absorption by H and $H_2$ in this region may be more challenging than previously thought. Based on their model of the Cassini/UVIS data, Koskinen et al. (2013) found that absorption in the Lyman continuum region arises mostly from the LW bands of $H_2$. As a result, they were only able to retrieve upper limits on the abundance of H. These upper limits, however, are generally less than 5 % below the exobase and thus consistent with the Voyager/UVS results from both Smith et al. (1983) and Vervack and Moses (2015). This means that the thermosphere of Saturn is dominated by $H_2$ at all altitudes below the exobase (~ 0.5–1 picobar).

### 9.2.2 Cassini UVIS occultation data

The Cassini/UVIS data represent a significant improvement and constitute the most extensive and reliable dataset to probe the upper atmosphere of any giant planet. Compared to the Voyager/UVS occultations, the Cassini/UVIS occultations afford higher S/N and a better point source resolution of 0.2 – 0.3 nm. The pointing stability of the Cassini/UVIS occultations is also excellent, thereby avoiding many of the issues that affected the Voyager observations. Finally, progress in computing, inversion techniques, and retrieval algorithms have undergone substantial evolution since the Voyager era (e.g., Koskinen et al. 2011, 2013, 2015; Vervack et al., 2004; Vervack and Moses 2015). This helps to improve the accuracy of the results, and to explore the reasons for the discrepancies in past analyses.

Since the orbit insertion in 2004, Cassini/UVIS has observed more than 40 stellar and solar occultations so far. Shemansky and Liu (2012) published results based on 3 of the stellar occultations and Koskinen et al. (2015) have recently analyzed 26 of the stellar occultations that were observed between 2005 and 2014 and provide useful information on the thermosphere. Koskinen et al. (2013) also analyzed 15 of the solar occultations to retrieve the density profiles and exospheric temperatures near the exobase of Saturn. The results of this study that were obtained by analyzing the ionization continuum of $H_2$ above the 0.1 nbar level (see Section 2.1) indicate that the exobase of Saturn is generally located between 2700 km and 3000 km above the 1 bar level and the composition of the thermosphere is dominated by $H_2$ with negligible dissociation and production of H at the probed latitudes.

Figure 9.2 shows the 41 density profiles of $H_2$ retrieved by Koskinen et al. (2013, 2015) from the solar and stellar occultations. The finite size of the Sun in the solar occultations means that the relationship between the observed transmission and altitude in the atmosphere is ambiguous. Thus the density profiles had to be parameterized and fitted to the data by using a forward model that accounts for the perceived size of the solar disk in the atmosphere (Koskinen et al. 2013). Stars are point sources and a combination of direct retrieval and forward modeling was used to analyze stellar occultations. In direct retrieval each spectrum is fitted separately to obtain the column densities of the absorbers as a function of tangent altitude. The column density profiles can then be inverted to retrieve number densities (Koskinen et al. 2011, 2015). At higher altitudes (z > 1800 km), however, transmission is close to unity and the retrieved column density profiles are too noisy for inversion. Thus the best fit forward model number density profiles are shown at high altitudes in Figure 9.2. Statistical simulations show that the density profiles are accurate to within about 5 – 20 %.

A comparison of the Voyager/UVS and Cassini/UVIS density profiles can be used to further test the validity of the results and detect differences in atmospheric properties between the Voyager and Cassini eras. Figure 9.3 compares four Voyager/UVS density profiles with Cassini/UVIS stellar occultation profiles at equivalent planetocentric latitudes. In general, the Voyager/UVS results agree well with the Cassini/UVIS results. Interestingly, the Cassini/UVIS densities at 21.1N agree well with the nearly symmetric V2 stellar ingress occultation at 21.7S while the Cassini/UVIS densities at 28.5N also agree well with the V2 solar egress occultation at 29N. This comparison puts upper limits on hemispheric differences near the equator and their time evolution between the Voyager and Cassini observations. The uncertainties in the Voyager/UVS density profiles, however, are often significantly larger than the uncertainties in the Cassini/UVIS profiles and they can mask temporal and spatial variations that we discuss in the next section. As argued by Vervack and Moses (2015), these uncertainties in the density profiles, have more than likely, also contributed to the ambiguities in the temperature retrievals (see Section 3.2).

The overall density structure in Figure 9.2 illustrates the fact that surfaces of constant pressure on Saturn can be approximated as deformed ellipsoids of revolution (e.g., Zharkov and Trubitsyn 1970). Thus the density profiles reach minimum radial distances at the poles and a maximum near the equator. In order to see if the isobars in the thermosphere are similar to the lower atmosphere, we plotted the radial distances of the 0.01 nbar level (slightly below the exobase of Saturn) as a function of planetocentric latitude based on the Cassini/UVIS and Voyager/UVS occultations in Figure 9.4. In both cases we retrieved the radial distances from the forward model density profiles (Koskinen et al. 2015; Vervack and Moses 2015). We note that the Voyager/UVS data are generally noisier and the associated uncertainty in the pressure levels is likely to be larger than the uncertainty in the Cassini/UVIS results. Thus we assigned an uncertainty of 50 km to the Cassini/UVIS pressure levels and an uncertainty of 100 km to the Voyager/UVS occultations.

The upper panel of Figure 9.4 indicates that the occultation results are broadly consistent with the expected shape of Saturn's atmosphere. A closer inspection, however, reveals that there are differences between the shape of the thermosphere and the lower atmosphere. The deviations of the thermosphere from the lower atmosphere are shown by the lower panel of Figure 9.4 that compares the 0.01 nbar level based on the occultations with the 100 mbar reference model of Anderson and Schubert (2007). Here we extrapolated the reference model to 0.01 nbar by adjusting the equatorial radius of the model to match three of the equatorial stellar occultations at planetocentric latitudes of 2N and 3S from late 2008 and early 2009. We note that there is a lively debate on the rotation rate of Saturn (see Chapter 5) that affects the wind-driven perturbations to the reference model. We chose the model of Anderson and Schubert (2007) for convenience because it minimizes the wind-driven perturbations while still matching the Voyager and Cassini gravity field parameters and the observed shape of the atmosphere (Lindal et al. 1985; Jacobson et al. 2006).

The deviations of the isobars in the thermosphere from the predicted shape (hereafter, the normalized altitudes) in Figure 4 show two interesting trends. First, the solar occultations (purple diamonds) from Cassini/UVIS indicate that the normalized altitude along the terminator limb increases with latitude away from the equator. The solar occultations in the southern hemisphere were all obtained between late 2007 and early 2008 so that they should be relatively free of time-dependent trends. The same is true of the solar occultations in the northern hemisphere that were obtained in 2010 with only two exceptions (the high latitude data point from 2007 and one of the low latitude data points from 2008). The trend of increasing normalized altitude with latitude was noted by Koskinen et al. (2013) who explained it by arguing that the thermosphere extends to deeper pressure levels at higher latitudes. The second trend is a relatively large 600 – 700 km scatter of the data points at low to mid (northern) latitudes in the lower panel of Figure 9.4. To our surprise this scatter appears not to be random. Instead, Figure 9.5 indicates that the exobase on Saturn expanded by about 500 km between 2006 and 2011, apparently followed by the onset of contraction some time after 2011. This trend is thought to arise from changes in the energy balance near the homopause that have caused the thermosphere to warm by about 100--200 K during the same time period (Koskinen et al. 2015), probably followed by cooling during contraction.

The lower panel of Figure 9.4 shows that the V2 results agree well with the Cassini/UVIS results from 2008/2009.  Within their uncertainty, they also agree with the time-dependent trend in Figure 9.5.  The V1 data points in Figure 9.4, however, appear significantly more elevated than the V2 data points.  This result may be compromised by the relatively large uncertainties in the V1 data and we do not give it much significance.  Overall, then, the Voyager/UVS points are consistent with the elevated state of the atmosphere that we observe in the Cassini/UVIS occultations of 2008/2009.  We note that the expansion and warming of the atmosphere in the Cassini/UVIS occultations anti-correlates with solar activity between 2006 and 2010.  Also, the Voyager/UVS and Cassini/UVIS observations of 2008/2009 occurred at opposite solar activity levels, indicating that the changes in the atmosphere are not driven by changes in solar activity.  Instead, the Voyager and Cassini observations coincided with the same season during the northern spring.  This implies that the observed changes, that are likely to arise from changes in dynamics, could be seasonal in nature.

### 9.3. REVIEW of RELEVANT TEMPERATURE DATA

#### 9.3.1. Molecular $H_3^+$ near-IR thermal emission

$H_3^+$ thermal emission has only been detected repeatedly in hotter, auroral/polar regions, where O'Donoghue, et al. (2014) report average thermospheric temperatures: 527±18 K in northern spring and 583±13 K in southern autumn seasons, respectively (see Chapter 7).  However, on different Saturnian days, the southern aurora has exhibited a much wider range of temperatures, varying between ~400-600 K (Melin et al., 2007; Stallard et al., 2012; Lamy et al., 2013). In contrast to Jupiter, $H_3^+$ emissions from the disk were detected only recently by O'Donoghue et al. (2013) and they are of intermittent nature as described in Chapter 8.  They are also not of sufficient quality to allow for the retrieval of temperatures.  Thus for low and mid latitudes one must rely on solar and stellar occultation measurements to infer temperatures in Saturn's thermosphere from the derived $H_2$ density scale heights.  Radio occultation data yield electron density profiles from which plasma scale heights can be derived, but there is no assurance that the electron, ion, and neutral gases are in thermal equilibrium so at most only the sum of electron and ion temperatures can be inferred (see Chapter 8 for plasma temperatures).

### 9.3.2. Inferred temperatures from H2 density profiles.

In line with the $H_2$ density profiles (see Figures 9.2 and 9.3), the retrieval of temperatures from the occultations is limited to pressures lower than 10 nbar. As we pointed out in Section 2.1, there have been significant disagreements over the exospheric temperatures retrieved from the occultations. For example, Broadfoot et al. (1981) used the V1 solar egress occultation at the planetocentric latitude of 30S to derive an exospheric temperature of 850 ± 100 K. Only a year later, Sandel et al. (1982) analyzed the V2 stellar egress occultation and reported an exospheric temperature of only 400 K at the planetocentric latitude of 3.5N. This result was contradicted by Festou and Atreya (1982) who retrieved a temperature of 800 K from the same stellar occultation.

Sandel et al. (1982) argued that the results of Broadfoot et al. (1981) may have been biased towards larger scale height and temperature by the finite size of the Sun that was not taken into account in the analysis. The angular size of the Sun at Saturn is about 1 mrad that, depending on the distance of the spacecraft from Saturn during the occultations, can translate to an apparent diameter of the solar disk comparable to or larger than the scale height of about 150 – 200 km of the thermosphere. Ignoring this effect, however, only overestimates the temperature by about 70 K even if the apparent diameter of the solar disk is as large as 500 km (Koskinen et al. 2013). Therefore we consider it unlikely that ignoring the solar disk in the analysis can explain a discrepancy of 400 K in the temperatures.

Instead, Smith et al. (1983) argued that the temperature retrieved by Broadfoot et al. (1981) was too large because the latter misinterpreted the effects of an instrument gain change during the occultation and had problems in dealing with severe pointing drifts. Vervack and Moses (2015), on the other hand, found that the high altitude light curves of the V1 solar egress occultation are actually consistent with a temperature of 800 K but also concluded that the high altitude results were corrupted by bad data. Thus both Smith et al. (1983) and Vervack and Moses (2015) support the conclusion that the relatively high temperature of 800 K is erroneous.

This does not explain the disagreement between Sandel et al. (1982) and Festou and Atreya (1982) over the V2 stellar egress occultation. In our opinion, however, Smith et al. (1983) already convincingly demonstrated that an exospheric temperature of 800 K does not provide as good a fit to this occultation as a temperature of 400 K.  They also derived an exospheric temperature of only 450 K at the planetocentric latitude of 29N from the ionization continuum of $H_2$ in the V2 solar occultation, which provides a fundamentally more robust measurement of the temperature than the the LW bands of $H_2$ that were used by Festou and Atreya (1982).  In addition, the lower temperature of 400 – 500 K is supported by the recent re-analyses of the V2 stellar egress occultation (Shemansky and Liu 2012; Vervack and Moses 2015).

   The lower temperature is also supported by the Cassini/UVIS occultations.  To show this, Figure 9.6 compares the exospheric temperatures from Cassini/UVIS (Koskinen et al. 2013, 2015) with the Voyager/UVS results (Smith et al. 1983; Vervack and Moses 2015).  The temperatures from Cassini/UVIS range from 370 K to 590 K, and the solar occultations in particular also indicate that the temperature increases by 100 – 150 K with latitude from the equator towards the poles.  In general, the Cassini/UVIS results are in good agreement with the Voyager/UVS data, with the exception of the V1 solar ingress occultation near the south pole at the planetocentric latitude of 84S.  This occultation, however, suffered from spacecraft slewing and anomalous channel behavior and Vervack and Moses (2015) do not place much significance on the disagreement with the Cassini/UVIS results there.  In our opinion the results from Cassini/UVIS, together with the re-analysis of the Voyager/UVS data, finally settle the debate on Saturn's exospheric temperatures.

   In addition to the exospheric temperatures, the occultations can be used to retrieve temperature profiles that are critical to our understanding of the energy balance in the thermosphere.  In general, there are three methods that have been used to retrieve the temperature profiles in the past.  First, a parameterized temperature profile can be fitted to the data by forward modeling the light curves or the retrieved column density profiles.  Second, the retrieved density profiles can be integrated directly to obtain partial pressures of $H_2$ that can be converted to temperatures by using the ideal gas law.  Third, the transmission spectra can be used to infer the rotational temperature of the $H_2$ molecules by analyzing the absorption bands.

Many of the past studies relied on forward modeling to estimate the temperatures (Festou and Atreya 1982; Smith et al. 1983). This approach can be dangerous, particularly if the uncertainty in the light curves and thus the density profiles is large (Vervack and Moses 2015). It is especially dangerous if the atmosphere models start from the 1 bar level because in that case the results depend on several free parameters that are not constrained by the data. Direct retrieval of temperatures has the advantage that it does not make any assumptions about the temperature profile and the uncertainties are tractable with Monte Carlo techniques (e.g., Koskinen et al. 2015). With large uncertainties in the density profiles, however, direct retrieval can introduce artificial waves to the temperature profiles and the exospheric temperature depends on the upper boundary pressure that is often not known *a priori*.

Both forward modeling and direct retrieval depend on the assumption of hydrostatic equilibrium. This is in general an excellent approximation in the thermosphere, but in principle it is also possible to constrain the rotational temperature of $H_2$ directly from the observed spectra. This approach was attempted by Shemansky and Liu (2012) who used a combination of forward modeling and spectral analysis to constrain the temperatures. They, however, concluded that the existing databases of $H_2$ absorption probabilities are not sufficiently extensive to reliably measure the temperatures. We note that spectral measurements of the temperature are also compromised by the insufficient wavelength resolution and S/N of the data. In addition, the absorption bands are affected by changes in temperature and level populations of $H_2$ along the line of sight that are not separable in the transmission spectra.

In order to reduce the associated uncertainties, Koskinen et al.(2015) used a combination of forward modeling and direct retrieval to obtain temperature and density profiles iteratively from the Cassini/UVIS occultations. For example, Figure 9.7 shows the temperature-pressure (T-P) profile based on an occultation of β Crucis from January 2009 that probes the atmosphere at the planetocentric latitude of 3S (hereafter, ST32). In this case the exospheric temperature is 427 ± 11 K and the uncertainty along the profile ranges from a few percent to about 15 %. Here the forward model profile agrees well with the direct retrieval. The uncertainty depends on the brightness of the star and altitude sampling rate of the occultations, and in this regard ST32 is one of the best datasets.

Curiously, the temperatures retrieved by Koskinen et al.(2015) disagree significantly with Shemansky and Liu (2012) for two of the three stellar occultations that were analyzed by the latter, i.e., ST32 and an occultation of $\delta$ Orionis from April 2005 that probes the atmosphere at the planetocentric latitude of 42S (hereafter, ST1). For ST1 Shemansky and Liu (2012) obtained an exospheric temperature of 318 ± 5 K whereas Koskinen et al.(2015) obtained a temperature of 429 ± 28 K for the same occultation. In addition, Shemansky and Liu (2012) obtained an exospheric temperature of 612 K for ST32 (cf. Figure 9.7).

Shemansky and Liu (2012) suggested that the scale height of $H_2$ decreases with altitude above 1400 km (1 nbar) due to significant dissociation of $H_2$. We note that the stellar occultations cannot be used to directly retrieve the abundance of H, and the idea that $H_2$ is significantly dissociated contradicts the relatively low abundances of H below the exobase that have been retrieved from solar occultations (Koskinen et al., 2013; Vervack and Moses 2015). Furthermore, Koskinen et al. (2015) did not find evidence for the dissociation of $H_2$ in the light curves from the stellar occultations that are actually consistent with the scale height increasing with altitude as expected. This suggests that dissociation of $H_2$ is not particularly important below the exobase and the $H_2$ density profiles are likely to be close to diffusive equilibrium above the homopause. As a result, the retrieval of temperatures in the thermosphere should not be significantly affected by uncertainties in the composition.

Assuming that the scale height of $H_2$ is not a reliable measure of the temperature in the upper thermosphere, Shemansky and Liu (2012) derived the temperature for ST32 by 'using a polynomic fit to the scale height' near 1400 km in altitude. We note that this method is not accurate in regions where the temperature changes with altitude, and it is typically much less accurate than forward modeling the density profiles or direct retrieval of temperatures (see above). To highlight this point, Figure 9.7 shows a comparison between the temperature profiles retrieved by Koskinen et al. (2015) and Shemansky and Liu (2012) for ST32. It is difficult to believe that a heat conducting atmosphere can support the temperature profile retrieved by Shemansky and Liu (2012) where the temperature increases by about 500 K within practically a single pressure level.

Finally, the extended analysis of the stellar occultations by Koskinen et al. (2015) allows for a more systematic exploration of the temperature structure in the thermosphere. A particularly fruitful method to probe thermal structure in the atmosphere is to combine the T-P profiles from the occultations with Cassini/CIRS data to create T-P profiles that extend from the 1 bar level to the thermosphere. For example, Figure 9.8 shows five temperature profiles based on three occultations from the spring of 2006 (hereinafter, ST5, ST10 and ST12) and two from December 2008 (hereinafter, ST30 and ST31). ST5, ST10, ST12 and ST31 probe the atmosphere near the planetographic latitude of 20N while ST30 probes the atmosphere near 2N. These occultations were chosen because of a close coincidence between the UVIS occultations with the CIRS measurements in the spring of 2006, and for the fact that most of the data points showing the expansion of the atmosphere between 2006 and 2011 lie in this region. The figure also shows the best fit forward model mixing ratios of $CH_4$ for the occultations that are discussed further in Section 7.

The temperatures in the lower atmosphere are retrieved from $CH_4$ emissions in the Cassini/CIRS limb scans and the results are valid up to the 3 μbar level. As we pointed out before, the Cassini/UVIS retrievals are valid down to the 0.1 – 0.01 μbar level, depending on the occultations. The implied agreement between the CIRS and UVIS temperatures is relatively good and the data indicate that the location of the base of the thermosphere varies between 0.1 and 0.01 μbar. We note, however, that this region falls into a gap in coverage between the two instruments, and thus we are prevented from accurately locating the base of the thermosphere. This also introduces additional uncertainty to the hydrocarbon mixing ratio profiles that are derived from the occultations (see Section 7).

Interestingly, the temperature profiles from December 2008 in Figure 9.8 are generally hotter than the corresponding profiles from the spring of 2006 in the lower thermosphere (~ 0.1-10 nbar) while there are no detectable differences in the exospheric temperatures. The base of the thermosphere may also be at a higher pressure level in the December 2008 occultations. This supports the argument by Koskinen et al. (2015) that warming and extension of the thermosphere to deeper pressures can explain the expansion of the atmosphere and, by inference, that the contraction of the atmosphere that may have started after 2011 is accompanied by cooling of the lower thermosphere. The origin of these changes in thermal structure, however, is currently poorly understood. A more comprehensive study that uses photochemical and radiative transfer models to interpret the temperature profiles together with the hydrocarbon abundances can shed further light on these processes and may provide more detailed information on dynamics in the mesosphere and lower thermosphere.

## 9.4. REVIEW of AIRGLOW DATA

In common with the $H_2$/He atmospheres of Jupiter, Uranus, and Neptune, Saturn's airglow is dominated by $H_2$ electronic bands, the He 58.4 nm line, the H Lyman line series, and $H_3^+$ near-IR bands. Because each atmosphere has a thermosphere significantly hotter than would be predicted by solar EUV and FUV heating, other energy sources must be considered to understand the mechanisms for airglow emission. A discussion of airglow is further challenged by the long-term calibration issues in the EUV/FUV for space-borne spectrometers, the low spectral resolution of the Voyager Ultraviolet Spectrometers (UVS), and the "no resolution" of the Pioneer 10 photometer. Thus one looks to the Hopkins Ultraviolet Telescope (HUT), Hubble Space Telescope (HST), and Cassini /UVIS for high spectral resolution, well-calibrated data. Without accurate absolutely calibrated data, a discussion of airglow is reduced to purely qualitative statements without any firm understanding.

### 9.4.1 H Lyman Alpha

In principle, the H Lyman-$\alpha$ (121.6 nm) dayglow on Saturn should be quite straightforward to explain. The strong solar Lyman-$\alpha$ line, with a line width of ~ 0.1 nm characteristic of line formation in a region where the temperature is ~ $10^{4-5}$ K in the solar atmosphere, is resonantly scattered by Saturn's atomic hydrogen above the $CH_4$ absorbing region, whose upper bound is approximately the homopause. The thermospheric temperature, ~ 300 - 600 K, governs the intrinsic planetary line width and the H column density above the absorbing $CH_4$ region governs the scattering optical depth at line center, and together they determine what fraction of the solar line can be resonantly scattered out of the atmosphere. While the thermospheric scattering optical depth at line center can be very large, up to $10^5$, it may be optically thin in the wings of the solar line, due to the mismatch of line widths associated with the mismatch of line formation temperatures, ~ 400 K vs. ~ 30,000 K. In addition radiative transfer to properly compute planetary line formation and the emergent intensity from the atmosphere must include angle dependent scattering with frequency redistribution (cf. Lee and Meier, 1980).

Voyager UVS and Cassini UVIS observations yield a relatively flat center-to-limb variation (Ben-Jaffel et al., 1995; Gustin et al., 2010), which would suggest optically thin emission (e.g. Ben-Jaffel et al. 2007 and references therein), even though the Saturn line is optically thick at line center. If total emission were dominated by line center photons, then a conservatively scattering atmosphere would give a center-to-limb cosine-like variation. However as the limb is approached the optically thin wings of the Saturn line become a more important source of emission and produce a flatter center-to-limb variation.

The two Voyager UVSs measured Saturn's H Lyman-$\alpha$ brightnesses at ~ 3.3 kR (V1) and 3.0 kR (V2), (Broadfoot et al., 1981; Sandel et al. 1982). The average Lyman-$\alpha$ disk brightness from 29 IUE observations was 1.1 ± 0.36 kR (McGrath and Clarke 1992). This discrepancy between UVS and IUE is perplexing in light of their agreement on the Jovian Lyman-$\alpha$ brightness. Gustin et al. (2010) give peak limb brightness values with adjustments for solar activity for V1: 1.9, 2.5 kR; V2: 1.8 kR; to be compared with their UVIS limb scans with peak brightness of only 0.8 kR and scan averages of 0.44 kR. But in Table 3 of Shemansky et al. (2009) the UVIS non-auroral Lyman-$\alpha$ brightness range near the limb is stated to be higher at ~ 1-1.2 kR, with reference to Shemansky and Ajello (1983) that the V1 brightness was larger, ~ 4.9 kR, at mid-latitudes in 1980. No UVIS center of the disk nor disk averaged values have been published to facilitate a better comparison, but the prudent conclusion would be that the Voyager values need downward adjustment.

### 9.4.2 He 58.4 nm

Like Lyman-$\alpha$, the interpretation of the He I 58.4 nm line should also be straightforward were it not for the requirement that the He/$H_2$ ratio be accurately known. Planetary He absorbs solar He I 58.4 nm radiation and reemits/scatters it with a probability equal to 0.9989. In addition, knowledge of the thermospheric temperature for planetary line width, and location of the homopause for the He column density above the unit $H_2$ absorption optical depth are necessary for accurate interpretation of He I 58.4 nm observations and all are uncertain to various degrees.

The originally reported Voyager brightnesses were V1: 2.2 ± 0.3, and V2: 4.2 ± 0.5 R (Broadfoot et al., 1981; Sandel et al. 1982), whereas Parkinson et al. (1998) reported these measurements as disk center brightness values of V1: 3.1 ± 0.4 and V2: 4.2 ± 0.5 R, with no discussion for the increased V1 value.

Parkinson (2002) performed the most recent analysis of the Saturnian He 58.4 nm line brightness for Voyager UVS, for which some aspects were previously reported in Parkinson et al. (1998). Constrained by the Voyager IRIS He/H2 mixing ratio ~ 0.03 and UVS occultation data, Parkinson (2002) required an implausibly high homopause altitude and large vertical mixing of $K_{zz} > 10^9$ cm$^2$ s$^{-1}$, whereas if a solar He/H2 mixing ratio ~ 0.13 were appropriate as Conrath and Gautier (2000) inferred from reanalysis of IRIS data, then $K_{zz} > 2 \times 10^7$ cm$^2$ s$^{-1}$ for V1 and $K_{zz} > 1 \times 10^8$ cm$^2$ s$^{-1}$ for V2, with the latter still exceedingly large. It must be kept in mind that the Voyager He 58.4 nm brightnesses might need downward adjustment.

### 9.4.3 H$_2$ Electronic Bands

The surprisingly large H$_2$ EUV/FUV dayglow intensities observed by Voyager for Jupiter, Saturn, and Uranus generated a lively debate about excitation mechanism(s), primarily because at the time there were no rigorous calculations available on strong solar line contributions to H$_2$ fluorescence in the dayglow. Three principal mechanisms were advanced to explain the dayglow: 1) additional electron excitation (Shemansky 1985), 2) dynamo-plasma acceleration (Clarke et al. 1987), and 3) solar fluorescence (Yelle 1988), in addition to dayglow generated by photoelectrons (cf. Strobel et al. 1991). The "excess" dayglow was given a name "electroglow" (Broadfoot et al. 1986), yet the measured intensities exhibited a dependence on the incident solar EUV and FUV fluxes at each planet. Broadfoot et al. (1986) emphasized excitation by low-energy electrons as a necessary component of the phenomenon. However, the power requirements to energize these electrons exceeded substantially what the Sun could supply in the UV from known processes.

It was the combination of the high resolution HUT spectra (0.3 nm) of Jupiter's dayglow (Feldman et al., 1993) and the definitive calculation performed by Liu and Dalgarno (1996), who demonstrated that solar-induced H$_2$ fluorescence creates a spectrum distinctly different from photoelectron impact on H$_2$ that explains Jupiter's dayglow.

The strongest fluorescence, ~ 14% of the total, is due to the solar Lyman-β line at 102.572 nm (as proposed by Yelle, 1988), which is coincident with the P(1) line of the $H_2$ Lyman 6-0 band at 102.593 nm.

Fortunately for Saturn Cassini UVIS data has a spectral resolution of ~ 0.55 nm, sufficient to separate the solar fluorescence contribution from electron impact generated $H_2$ band emissions. Gustin et al. (2010) used UVIS limb scan data taken at low latitudes below the ring plane to derive volume emission rates for various components of the dayglow. With disk-averaged Jupiter dayglow contributions adopted from Liu and Dalgarno (1996) adjusted for solar activity and scaled to Saturn, Gustin et al. (2010) obtained 173 R for fluorescence generated dayglow and 131 R for electron impact produced dayglow; thus in the ratio of 0.57:0.43. From the UVIS limb data, Gustin et al. (2010) derived limb-averaged values of 460 and 1054 R, respectively, with a ratio of 0.3:0.7. They noted that this ratio reaches a minimum of 0.2:0.8 at a tangent altitude of 1400 km which suggests that solar fluorescence is relatively more important on disk and relatively unimportant on the limb and that electron impact becomes progressively more important at high altitudes. A detailed analysis of disk-center dayglow would be extremely enlightening to determine whether solar fluorescence plus photoelectron-generated $H_2$ dayglow is sufficient to explain Saturn's dayglow as it is for Jupiter's dayglow.

### 9.4.4 $H_3^+$ Thermal Emission

The $H_3^+$ ion plays a fundamental role as a thermospheric thermostat for the giant planets in a manner analogous to NO in the Earth's thermosphere. By near-IR thermal emissions in its $\nu 2$ band, between 3.4–4.1 microns (described in detail in Chapter 7) $H_3^+$ regulates Saturn's thermospheric temperature. Saturn's low and mid-latitude thermosphere is colder (~ 400-450 K) with fewer $H_3^+$ ions. Thus $H_3^+$ thermal emission has mostly been detected in hotter (> 500 K) auroral/polar regions.

### 9.5. REVIEW OF COMPOSITION

Saturn's thermosphere is mostly $H_2$, with an uncertain amount of He and a maximum volume mixing ratio of H atoms at the exobase of 0.05, (Koskinen et al., 2013) and proportionally decreasing with decreasing altitude given the 2:1 ratio in $H:H_2$ scale heights above the homopause. We note that the UVIS occultation forward models (Section 9.7) are also more consistent with the higher mixing ratio of 0.13 from Conrath and Gautier (2000) while the lower bound of 0.03 leads to atmospheric structure that provides a worse fit to the $H_2$ and $CH_4$ density profiles retrieved from the occultations. From the He 58.4 nm line emission analysis by Parkinson et al. (1998) and Parkinson (2002), our inferred location of Saturn's homopause from UVIS occultation data, and downward revision of the Voyager 58.4 nm brightnesses, only a He/ $H_2$ ratio close to Jupiter's ratio of 0.157 could yield 58.4 nm intensities in the revised Voyager range.

To date there are no measurements of HD in the thermosphere, but it may be possible to detect HD with the Cassini Ion Neutral Mass Spectrometer (INMS) during the Proximal Mission when the Cassini spacecraft flies through Saturn's thermosphere. In Saturn's well-mixed lower atmosphere, there are a number of measurements of the D/H ratio in molecular hydrogen and methane, i.e., of HD and $CH_3D$. From the review of these measurements by Fouchet et al. (2009), one concludes that the HD/H2 ratio is ~ 3.5 x $10^{-5}$ with error bars of ~ ± 50%.

With a D/H ratio (= ½ HD/$H_2$) in the well-mixed atmosphere, H atoms with an upper limit of 5% at the exobase, and D with the same scale of height as $H_2$, the atomic D mixing ratio will be in the range of $10^{-8}$ to $10^{-7}$ (Parkinson et al., 2006), and hence of limited interest in the thermosphere.

For the purposes of this chapter the only real importance of $CH_4$, with a volume mixing ratio in the lower atmosphere of 0.0047 is to locate the homopause. Chapter 10 discusses $CH_4$ and its photochemistry in depth. Likewise, $H_2O$ is another minor species in the thermosphere, which is of much greater interest for Saturn's ionosphere, in connection with a phenomenon known as "ring rain" and discussed in depth in Chapter 8. $H_2O$ molecules are heavier than $H_2$ and have a large loss rate in the lower stratosphere due to chemical loss, if they survive condensation, as they diffuse downward through the atmosphere. If they diffuse at their maximum velocity, their volume mixing ratio, μ, is approximately the downward flux, φ($H_2O$), multiplied by the $H_2O$ scale height divided by the $H_2O$-$H_2$ binary collision coefficient and in cgs units μ($H_2O$) = φ($H_2O$)/$10^{13}$, essentially the "inverse Hunten limiting diffusive flux" for heavy gases (Hunten, 1973, cf. his Eq. 15). Thus, for example, a flux of 1 x $10^6$ $H_2O$ cm$^{-2}$ s$^{-1}$ estimated by Müller-Wodarg et al. (2012) near the planetocentric latitude of 20N yields a thermospheric mixing ratio of 1 x $10^{-7}$. Figure 9.9 illustrates density profiles and volume mixing ratios representative of the above discussion.

## 9.6. Inferred Net Heating Rate from Radial Temperature Profile

In the thermosphere molecular heat conduction is an important process for redistribution of thermal energy. A temperature profile yields from its gradient the heat conduction flux, $F_H = -\kappa \nabla T$, where $\kappa = 252\, T^{0.751}$ in ergs cm$^{-1}$ s$^{-1}$ K$^{-1}$ (Hanley et al., 1970) for a H$_2$ dominated atmosphere, and from its curvature the heating/cooling rate $\nabla \cdot F_H$. Occultation data yield fundamentally the line of sight (los) column density. The local number density is derived by inverting the column density profile and the temperature is inferred from the retrieved density profile. The partial pressures are first obtained by integrating the equation of hydrostatic equilibrium downward, starting from an assumed temperature and thus pressure at the upper boundary of the observed density profile. The ideal gas law then yields the temperature at each altitude point based on the derived pressure and the observed densities. An alternate approach is to create a model atmosphere with the temperature lower boundary condition from CIRS and wavelength dependent light curves that match the observed UVIS light curves. A comparison of both approaches is shown in Fig. 9.8 for UVIS stellar occultations obtained in 2006 and 2008, with diamonds for the data-only method and solid lines for the forward model approach.

The radial heat conduction equation, with r for radial distance, is: $-\frac{1}{r^2}\frac{\partial}{\partial r}\left(r^2 \kappa \frac{\partial T}{\partial r}\right) = Q(r) - C(r)$, where Q(r) and C(r) are the heating and cooling rates, respectively, and can be due to dynamical as well as radiative processes. If one transforms the heat equation from variable r to 1/u, an analytic solution is obtained in terms of Gaussian-like functions for Q(r) and C(r))(cf. Stevens et al., 1993). Derived heating and cooling rates are most valid if their sources are spatially well separated and the temperature profile being modeled is well constrained by data over the entire profile. Referring to Figure 9.8, CIRS data constrain temperature profiles reliably up to 3 μbar and can be extrapolated to 0.01 μbar. Only in exceptional circumstances such as the December 2008 do stellar occultations yield an adequate temperature profile in this critical region.

In Figure 9.10 an illustrative solution to the above heat conduction equation is given for December 2008 stellar occulation derived temperature profiles shown in Figure 8. Solution of the heat conduction equation yields a net integrated heating rate of 0.072 ergs cm$^{-2}$ s$^{-1}$, with peak heating at 1450 km and 0.65 nbar, while the peak cooling is inferred at 870 km and 70 nbar.

The 2008 occultation was at low latitude, 18 N, where the asymptotic temperatures are ~ 400 K, whereas at auroral/polar latitudes the temperature rises to values of 550-600 K. Thus a solution of the heat equation at high latitudes would yield a larger heating rate. If one performed a series of solutions at discrete latitudes and then globally averaged the rates, one would find a globally average heating rate of ~ 0.1 erg cm$^{-2}$ s$^{-1}$ or ~ 5 TW total for Saturn's thermosphere, in considerable excess of what solar EUV/FUV power can deliver (~ 0.15-0.3 TW). Note that this is the global heating rate and the required power input is the heating rate divided by the heating efficiency, which according to Waite et al. (1983) is ~ 0.5 for solar UV heating and auroral energy sources. Thus the power input required is ~ 10 TW, for which only Joule/ion-neutral heating can supply this amount as discussed in Section 8.3.

### 9.7. Inferred Homopause Location from CH4 Data

Absorption by $H_2$ in the occultations is negligible at wavelengths higher than about 120 nm and this allows for minor species such as $CH_4$, $C_2H_6$, $C_2H_4$, and $C_2H_2$ to be detected. The $CH_4$ profiles can then be used to constrain the eddy mixing coefficient $K_{zz}$ and the location of the homopause. Unfortunately mixing ratios are required to properly pinpoint the location of the homopause and absorption by $H_2$ is saturated at the level in the atmosphere where methane densities are retrievable (i.e., roughly below the 0.01 μbar level), making it difficult to determine the mixing ratio without interpolating between regions. Combining temperature measurements in the stratosphere with the temperature profiles from the UV occultations is therefore critical for creating atmosphere models that can be used to calculate the mixing ratios of $CH_4$ and other hydrocarbons (cf. Figure 9.9). The wealth of observations from Cassini/CIRS makes this approach more reliable for the Cassini/UVIS data than for the Voyager/UVS occultations.

The results are still subject to uncertainties, however, because CIRS and UVIS do not observe the same location at the same time, and there is a gap in the temperature coverage of the two instruments between 0.01 µbar and 3 µbar (see Section 3.2).

With these caveats in mind, Vervack and Moses (2015) draw two conclusions based on the Voyager/UVS occultations that can now be re-evaluated in light of the Cassini data. First, Saturn's upper atmosphere is subject to strong mixing with a relatively high altitude homopause and second, the location of the homopause may be highly variable. Based on their analysis of five Voyager/UVS occultations, Vervack and Moses (2015) found that the $CH_4$ profiles could only be fitted by $K_{zz}$ profiles that increase with altitude throughout the thermosphere. Thus the homopause pressure, where by definition $K_{zz}$ is equal to the $CH_4$ – $H_2$ molecular diffusion coefficient, was typically very low. The Voyager 2 solar ingress occultation near the planetocentric latitude of 29N showed the lowest pressure homopause at 0.7 nbar with $K_{zz} = 2 \times 10^9$ $cm^2s^{-1}$. We note that such high values of $K_{zz}$ agree with the inference of strong mixing from the Voyager/UVS He 58.4 nm data (see Section 4.2).

Given the behavior of the $K_{zz}$ profiles in their atmosphere models, Vervack and Moses (2015) found it more convenient to derive the pressure and $K_{zz}$ at the level where the mixing ratio of $CH_4$ is $5 \times 10^{-5}$ (hereafter, the $CH_4$ reference level), than locating the homopause, to facilitate comparison with other work and to look for variations. In the three Voyager/UVS occultations probing the southern hemisphere the $CH_4$ reference level was located at 0.01 – 0.1 µbar with $K_{zz} = 1 - 3 \times 10^7$ $cm^2s^{-1}$. In the two occultations probing the northern hemisphere, on the other hand, the $CH_4$ reference level was closer to 0.01 µbar with a higher $K_{zz}$ of $(1-2) \times 10^8$ $cm^2s^{-1}$.

Similar background atmosphere models based on the Cassini/UVIS observations have only been developed for five occultations to date. The model temperature profiles and the resulting $CH_4$ mixing ratios are shown in Figure 9.8. Koskinen et al.(2015) retrieved these $CH_4$ profiles from the FUV channel of the Cassini/UVIS instrument and created the atmosphere models. A more comprehensive analysis of all of the Cassini/UVIS occultations is in progress and it will provide highly anticipated global constraints on the variability of the homopause and associated dynamics.

Meanwhile, the results from Cassini/UVIS so far are generally more consistent than from the Voyager data,

particularly because they do not confirm the peculiarly low pressure homopause in the Voyager 2 solar ingress occultation. Four of the occultations in Figure 8 probe almost the same location as the Voyager 2 solar occultation near the planetographic latitude of 20N and they indicate that the homopause pressure is 0.01 – 0.1 µbar with $K_{zz}$ = $10^6$ – $10^7$ $cm^2 s^{-1}$. The $CH_4$ reference level based on these occultations, on the other hand, is located closer to the 0.1 µbar level with similar values of $K_{zz}$ as at the homopause. These results agree reasonably well with the Voyager/UVS results in the southern hemisphere but not in the northern hemisphere.

We note that the Cassini fits to the $CH_4$ profiles are in agreement with the Voyager/UVS results in that the $K_{zz}$ profiles that are required to match the data often increase with altitude until relatively low pressures. This differs from the typical behavior of the $K_{zz}$ profiles in many planetary atmosphere applications that are assumed to asymptote to a constant value at some point in the thermosphere. The curious behavior of the $K_{zz}$ profiles could arise from photochemical processes that, contrary to expectations, affect the $CH_4$ profile, and/or waves or other dynamical processes that are not captured by the form of the $K_{zz}$ profile assumed in the current studies.

### 9.8. ENERGETICS OF THE THERMOSPHERE

#### 9.8.1 Inadequacy of Solar EUV/FUV Heating

Strobel and Smith (1973) reviewed the literature on calculations of the temperature of the Jovian thermosphere and performed new calculations for Jupiter, Saturn, and Titan. For Saturn, they estimated that solar EUV/FUV heating could raise the asymptotic isothermal thermospheric temperature by only ~ 10 K above the mesopause temperature. As noted in Section 6, the inferred heating rate from thermospheric temperature profiles far exceeds what the Sun can supply at EUV/FUV wavelengths.

#### 9.8.2 Wave Heating

The possibility of wave heating was evaluated in Strobel (2002) for the giant planets' thermospheres based on previous detailed calculations performed by Matcheva and Strobel (1999) for gravity waves in Jupiter's thermosphere. With appropriate values for the input parameters, dynamic viscosity, µ, gravitational acceleration, g, and gas constants, $c_P/R$, the maximum gravity wave energy flux in isothermal regions for Saturn is just $3.22\frac{\mu g R}{c_p}$ = 0.13 erg cm$^{-2}$ s$^{-1}$ (corrected expression from Strobel, 2002), which when coupled with the estimated heating efficiency, ~ 0.41, reduces the maximum integrated heating rate to ~ 0.055 erg cm$^{-2}$ s$^{-1}$, too low by about a factor of 2, and less if wave heating were not globally distributed and continuously active.  The latter conditions are extremely improbable.

   Another important class of vertically propagating internal waves is Rossby waves whose restoring force is the meridional variation of the Coriolis force and whose dynamics are based on conservation of potential vorticity.  Generally the potential vorticity of the atmosphere is dominated by planetary vorticity (f = twice the rotation rate times the sin(latitude)) with a minor contribution from the relative vorticity of the velocity field, $\nabla \times \vec{v}$.  As Rossby waves propagate vertically they must extract potential vorticity from the mean flow, $q_0$, in order for their wave potential vorticity, q', to grow exponentially in amplitude as $\rho_0^{-0.5}$, in the absence of dissipation.  But the wave potential vorticity cannot exceed the basic state potential vorticity, i.e., q' < f, and this restricts wave amplitudes to two orders of magnitude lower than estimated by amplitude growth (Schoeberl and Lindzen, 1982).

   The last class of propagating waves is acoustic waves, which are generated by lightning and thunderstorms (Schubert et al., 2003).  Their amplitudes and associated energy fluxes are poorly constrained.  To reach the thermosphere, their horizontal phase speeds need to be supersonic relative to the local speed of sound or otherwise they will be refracted by the thermosphere's increasing index of refraction.  This requires that the storms launching these waves must be moving at supersonic speeds in the troposphere (> 1.5 km s$^{-1}$) and three times the speed of the equatorial tropospheric jet.

### 9.8.3 Joule (Ion-Neutral) Heating due to Magnetosphere-Ionosphere-Atmosphere Coupling

In the upper atmosphere, the presence of ionospheric plasma provides a medium which responds to the presence of magnetic and electric fields, and thereby to processes that occur in the magnetosphere. When a magnetospheric electric field maps along the magnetic field lines into the atmosphere, the ionospheric ions are accelerated and collide with the ambient neutral gas particles. This collisional interaction leads to an acceleration of the neutral gases in the direction of the zonal ion drift, generating a region of large zonal wind velocities where magnetospheric electric fields are strongest, near the auroral emission regions. The acceleration is given by **a** = $-\nu_{ni}$(**u**-**v**) = (**j** x **B**)/$\rho$, $\nu_{ni}$ being the neutral-ion collision frequency, **u** the neutral wind vector, **v** the ion velocity vector, **B** the planetary magnetic field and $\rho$ the mass density of the neutral atmosphere. The electrical current density perpendicular to the magnetic field in the ionosphere is **j** = $\underline{\sigma}$(**E** + **u** x **B**), with $\underline{\sigma}$ being a tensor with components for the Pedersen conductivity and the Hall conductivity, **E** being the sum of the magnetospheric electric field mapped into the upper atmosphere and any polarization field set up by divergence of **j**. Because the ionosphere is not perfectly conducting, resistive heating occurs, a process often referred to as Joule heating. The thermal heating of the atmosphere by electrical currents per unit mass can be written as $q_{Joule}$ = (**j**·**E**)/$\rho$ where the electrical current density **j** and the electric field **E** both include the effect of neutral winds via the dynamo field term, **u** x **B** (Vasyliunas and Song 2005, equation 43). We note that electrical currents also result in momentum change due to ion drag that affects the kinetic energy of the gas. Sometimes this latter effect is referred to as "ion drag heating". While the thermal heating by currents alone (without considering neutral winds) can only be a positive quantity, the ion drag heating, $q_{Ion}$, can also attain negative values, implying the loss of kinetic energy of the neutral atmosphere. As a result, the calculation of ion drag heating requires knowledge of the thermospheric winds.

    The Saturn Thermosphere Ionosphere Model (STIM) is a General Circulation Model (GCM) which numerically solves non-linear coupled Navier-Stokes equations of energy, momentum and continuity for both neutral gas particle and ions in Saturn's thermosphere and ionosphere (Müller-Wodarg et al., 2006; 2012). The model currently relies on provision of magnetospheric electric fields and electron energy fluxes as external boundary conditions but then calculates the magnetosphere-ionosphere-thermosphere interaction self-consistently. The model includes solar and electron impact ionization, using for the latter the parameterization of Galand et al. (2011). Once created, the ions undergo chemical reactions as described by Moore et al. (2004).

Figure 9.11 shows the magnetospheric electric field strength (color contours) as mapped into the southern polar region. Also shown are the locations of maximum field-aligned current (black symbols), which coincide with the regions of largest electron precipitation into Saturn's polar upper atmosphere. The values of Figure 9.11 are taken from the BATSrUS MHD model of Saturn's magnetosphere (Jia et al., 2012) for quiet solar wind conditions. We have multiplied the original field strength by a factor of 4 in order to better reproduce observed polar temperatures and winds. The electric field is directed primarily equatorward and thereby generates a westward acceleration of the ionospheric ions.

Electron precipitation occurs along the ring-shaped region in Figure 9.11 and is local time dependent not only in terms of its latitude (as seen in the figure) but also in terms of the magnitude of the electron energy flux. We apply in STIM-GCM the local time shape of electron flux consistent with that inferred from auroral observations by Lamy et al. (2009) with a maximum flux in the dawn sector near 08:00 (Müller-Wodarg et al., 2012). Near midnight the electron flux is close to zero. In the simulation shown here we apply 10 keV electrons alone, but the model allows for implementation of other electron populations as well. We assume a longitudinally averaged auroral energy flux of 1.0 mW m$^{-2}$, a value based on the findings of Lamy et al. (2009). The electron impact ionization causes enhanced Pedersen and Hall conductivities in the atmosphere, which closely follow the local time changes of electron precipitation.

Figure 9.12 shows zonally averaged Pedersen conductances in Saturn's ionosphere as a function of latitude, assuming that magnetic field lines are aligned radially in the thermosphere. Since the auroral magnetospheric interaction is confined to polar latitudes, this assumption is acceptable in Saturn's almost perfect dipole field (see Chapter 4). At low latitudes the conductance results from solar radiation ionization, reaching around 2-3 mho, depending on the season (larger at equinox). These rapidly decrease towards the poles with increasing solar zenith angle. From 70-75° latitude, however, we see a strong enhancement to values of around 5-7 mho which result from the 10 keV electron impact ionization. The figure shows no seasonal variation of conductances at auroral latitudes but a hemispheric asymmetry. Southern auroral conductances may attain 7 mho, while those in the north reach around 5 mho only. This difference is a direct consequence of the magnetic field asymmetry between north and south. We assume the Saturn Pioneer Voyager (SPV) magnetic field model in our simulations (Davis and Smith, 1990).

Figure 9.13 shows neutral temperature contours and the meridional circulation wind vectors, as simulated by STIM for equinoctial conditions. The longest arrow corresponds to around 350 m s$^{-1}$. Also shown in the figure are two line plot panels with the normalized quantities: Joule heating rates (solid line), ion drag acceleration (dashed), and zonal wind velocities (dashed-dotted). The red dot in the temperature panel denotes the location of maximum Joule heating, the green dot denotes the region of maximum zonal ion drag. Both occur at 72°S latitude but around 100 km apart vertically. The curves on the right panel are vertical profiles at this latitude while the curves on the top panel are latitudinal profiles at the height of peak Joule heating (solid line) and at the height of peak ion drag (dashed and dashed-dotted lines). Peaks of Joule heating and ion drag in our simulation are slightly below the region of peak $H_3^+$ emission (1155±25 km) observed by Stallard et al. (2012). Zonal winds in the upper panel are normalized to a value of 334 m s$^{-1}$, and their largest values in the right panel reach 1500 m s$^{-1}$, the local sound speed. The peak values of zonal ion drag and Joule heating in the two panels are 0.02 m s$^{-2}$ and 1.9x10$^{-8}$ W m$^{-3}$, respectively.

As expected, the 72°S locations of maximum Joule heating and ion drag coincide with the region of peak electron precipitation and largest resulting conductance (see Figure 9.12). Interestingly, the largest zonal winds occur more poleward at 78°S (see top panel), showing the influence of pressure gradients and Coriolis acceleration as additional factors affecting the winds. A westward zonal wind will experience poleward Coriolis acceleration. While ion drag is largest in the deeper ionosphere near 900 km (green dot in main panel of Figure 9.13 and dashed line in right panel), zonal winds are essentially in a gradient wind balance aloft driven by poleward directed pressure gradients on equipotential surfaces generated by auroral heating and resulting in increasing temperatures with latitude as illustrated in Figure 9.13. Ground based Doppler analyses of $H_3^+$ emissions have revealed zonal ion velocities at polar latitudes on Saturn reaching supersonic speeds of several km s$^{-1}$ (Stallard et al., 2007), described in Chapter 7. As shown by Müller-Wodarg et al. (2012), plasma velocities for the conductances encountered in our simulations (Figure 9.12) can exceed neutral velocities by around a factor of 2, so our simulations are broadly consistent with these observations.

The temperatures, like zonal winds, are not largest in the region of strongest coupling with the magnetosphere (red and green dots) but instead peak in the polar cap region near 1300 km altitude, decrease again towards higher altitudes and reach their asymptotic values near 2000 km. Investigation of the energy equation terms in STIM (Müller-Wodarg et al., 2012) reveals this behavior to be a direct consequence of the transport terms, advection and adiabatic heating and cooling. As illustrated in Figure 9.13 (red arrows), the polar thermosphere hosts a complex circulation pattern in the meridional/altitude plane, with anti-clockwise and clockwise circulation cells, respectively, on the poleward and equatorward side of 72°S below 1500 km. The poleward flow below 1500 km transports energy away from the region of peak Joule heating and downwelling over the polar region causes adiabatic compression and heating near 1100-1600 km. The red dotted line in Figure 9.13 indicates the boundary between upward and downward winds. At higher altitudes, the circulation is broadly from pole to equator, but broken into several clockwise circulation cells due to Coriolis forces. This cools the polar region above 1600 km adiabatically (causing the negative temperature gradient there, as seen in Figure 9.13) and transporting some of the energy equatorward (thereby causing a slight temperature increase with altitude equatorward of 80°S).

These simulations illustrate that Saturn's high latitude thermosphere and ionosphere are driven primarily by coupling to its magnetosphere. The strong westward winds, which reduce the degree of corotation of the thermosphere to only 25% near 78°S, are a signature of angular momentum transfer from the upper atmosphere into the magnetosphere. The atmospheric response to this localized coupling to the magnetosphere spreads over the entire high latitude region poleward of 60° in both hemispheres. The zonal winds rapidly decrease towards more equatorial latitudes as a result of angular momentum conservation. Joule heating in our simulations provides about 6 TW of thermal energy into Saturn's upper atmosphere (summed over both hemispheres), around 20-40 times the total energy deposited by solar heating. This further emphasizes the importance of magnetosphere-atmosphere coupling on Saturn and giant planets.

### 9.8.4 Resistive Heating and Ion Drag by Wind-Driven Electrodynamics

Outside the auroral regions, the neutral atmosphere can be affected by resistive heating and ion drag driven by electric fields that arise from the interaction of the ionospheric plasma with neutral winds, turbulence or waves.  This interaction is known to be important in the E and F region of the Earth's ionosphere (e.g., Richmond et al. 1992; Richmond 1995; Richmond and Thayer 2000).  It should be important also in Saturn's ionosphere, given the importance of electrodynamics in the high latitude thermosphere, and could play a role in explaining the remarkable variability in the electron density profiles retrieved from Cassini/RSS observations (Nagy et al. 2006; Kliore et al. 2009, see Chapter 8).  It may also interfere with the circulation driven by auroral heating and ion drag, modify the ionospheric electric fields that are mapped down from the magnetosphere, and heat the non-auroral thermosphere.

   The generation of ionospheric currents in general relies on plasma-neutral collisions that force the electrons and ions to move at different velocities across magnetic field lines, thus violating the frozen-in flux condition of ideal magnetohydrodynamics.  Wind-driven electrodynamics or the ionospheric dynamo, on the other hand, is based on the generation of polarization electric fields by neutral winds that lie perpendicular to the magnetic field lines and, due to high field-aligned conductivity, remain approximately constant along magnetic field lines that traverse different layers of the atmosphere.  On the Earth, for example, the dayside dynamo layer is in the E region, and the electric fields are mapped between the E and F regions.  The result is an electric circuit that allows current to flow in the F region, leading to ion drag and resistive heating.  At night when the E region electron density diminishes, however, the polarization electric fields that are set up in the F region significantly reduce the current density.  One suggested system of thermospherically-driven currents at Saturn is a polar twin-cell vortex that has been evoked to drive oscillations in the planetary period, as described in Chapter 5.

   To further understand the ionospheric dynamo, it is convenient to divide Saturn's ionosphere into different "magnetization" regions M1, M2 and M3 (e.g., Koskinen et al., 2014).  In the M1 region (> 10 µbar) both the electrons and ions are collisionally coupled to the neutrals and currents are generally negligible.  In the M2 region (0.01-10 μbar), which is similar to the Earth's E region dynamo layer, the electron gyrofrequency is higher than the electron-neutral collision frequency while the ions remain collisionally coupled to the neutrals. In the M3 region (< 0.01 μbar) both the ion and electron gyrofrequencies are higher than the ion/electron-neutral collision frequencies. By definition, the Hall conductivity dominates in the M2 region while the Pedersen conductivity dominates in the M3 region.

Recently, Smith (2013) suggested a new source of thermospheric heating on Jupiter based on electrodynamic coupling of the thermosphere and stratosphere that relies on wind-driven electric fields. His study shows that in principle the ionospheric dynamo can generate perpendicular electric fields on the order of 10 mV m$^{-1}$ that result in resistive heating rates that are sufficiently large to explain the high temperatures in the Jovian thermosphere. For Saturn's equatorial thermosphere, the required peak heating rate to explain the low latitude temperature profiles retrieved from the Cassini/UVIS occultations is about 10$^{-10}$ W m$^{-3}$. Given that the peak Pedersen conductivity is about 10$^{-5}$ S m$^{-1}$ (Moore et al.2010), an electric field of about 3.2 mV m$^{-1}$ in the neutral reference frame would produce the required heating rate. The maximum dynamo electric field strength can be estimated as ≈UB where U is the wind speed, which implies winds of ~ 150 m s$^{-1}$ to generate the required electric field. While not impossible, these winds are needed in the M2 region (0.01-10 μbar) where we have no data.

Whether this mechanism actually turns out to be feasible on Saturn, however, depends on a number of assumptions. For example, Smith (2013) assumed zero winds in the thermosphere. This is problematic because the electric field in the neutral reference frame is given by $\mathbf{E_n}$ = $\mathbf{E}$ + $\mathbf{u}$ x $\mathbf{B}$ where $\mathbf{u}$ is the neutral wind and $\mathbf{E}$ is the dynamo electric field. The assumption of zero winds thus provides only a crude estimate of the current density that is likely to be an upper limit because the strength of the dynamo field also depends sensitively on the winds in both the M2 and M3 regions (Koskinen et al., 2014). For example, if the current initiated in the M3 region cannot close in the M2 region, a polarization electric field $\mathbf{E}$ is set up that cancels out the current in the M3 region. In addition, an electric field of 3.2 mV m$^{-1}$ at low latitudes would lead to substantial ion drag on the neutral atmosphere that affects the heating rates. Resistive heating and ion drag must therefore be modeled self-consistently by a circulation model that includes ionospheric electrodynamics.

Lastly, electrodynamic coupling and the ionospheric dynamo rely on substantial conductivities in both the M2 and M3 regions. The M2 region on Saturn lies almost entirely in the hydrocarbon ion layer where recombination rates are much faster than in the M3 region. Nevertheless, recent calculations by Kim et al. (2014) indicate that electron densities ~ $10^3$ cm$^{-3}$ are possible in the M2 region, but the complex ion chemistry makes the identity of major heavy ions, effective dissociation recombination rates, and calculated electron densities uncertain. While radio occultations do not rule out significant electron densities in the M2 region, this region is at the limit for retrieving reliable electron densities.

### 9.9. Global General Circulation Models of the Thermosphere

Apart from Doppler measurements of $H_3^+$ emissions at auroral latitudes, no observational evidence exists of the winds in Saturn's (or any giant planet's) mesosphere and thermosphere. We thereby rely on the use of numerical models to examine the general circulation of Saturn's mesosphere and thermosphere, for which only the STIM model has published results (Müller-Wodarg et al., 2006; 2012). But for Jupiter, however, several such models have been published (Achilleos et al, 1998; Bougher et al., 2005; Tao et al., 2014). These build on a heritage of thermosphere-ionosphere models for Earth which have been adapted to simulate the giant planet environment. GCMs numerically integrate the non-linear coupled Navier Stokes equations of momentum, energy and continuity on a global spherical grid. For giant planets the most common vertical coordinate used is pressure, based on the hydrostatic assumption for a high gravity environment.

The models require the inclusion of magnetosphere drivers, namely particle (mostly electron) precipitation and the magnetospheric convection electric fields. These appear as sources of ionization (alongside solar EUV), thermal energy and momentum in the codes. For Earth, detailed knowledge is available of the high latitude electric convection fields and exact locations of electron precipitation. For giant planets, locations of electron precipitation are thought to coincide with locations of peak auroral UV emissions and can be inferred geographically, along with the electron energy fluxes, from observations and modeling (cf. Chapter 7). The electric fields are more challenging to obtain as no in-situ measurements are available. Assuming the magnetosphere-ionosphere coupling processes via Birkeland currents as originally proposed for Jupiter by Hill (1979), observations of the degree of corotation in the magnetosphere plasma may in principle yield estimates of electric fields. For Saturn, the model of Müller-Wodarg et al. (2012) initially assumed high latitude electric fields based on magnetospheric plasma flow patterns calculated by Cowley et al. (2004) but most recently replaced these with electric fields calculated by the BATSrUS Saturn magnetosphere MHD model (Jia et al., 2012).

The other important aspect relates to the electron precipitation. Energetic electrons from the magnetosphere are known to induce auroral emissions on Saturn in the EUV and FUV as well as the IR (Kurth et al., 2009). Electron precipitation occurs primarily along a narrow auroral oval region located between 70° and 85° latitude of width 1.5°-3.5° (Badman et al., 2006). The electrons have a mean energy of around 10 keV but observations have identified energies ranging from 400 eV to 30 keV (Sandel et al., 1982; Gérard et al., 2004, 2009; Gustin et al., 2009). Including the effects of this electron precipitation on the ionosphere requires calculation of the collisional interaction between the electrons and atmosphere, including secondary ionization. This is best done via numerical solution of the Boltzmann equations for suprathermal electrons, which for practical reasons is done in 1-D rather than 3-D. Several such models have been proposed for Jupiter and Saturn (Grodent et al., 2001; Gustin et al., 2009; Galand et al., 2011) and STIM relies on the parameterization for Saturn which was developed by Galand et al. (2011) on the basis of full 1-D calculations. This parameterization provides vertical profiles of ionization rates for different electron populations, scaled by the background neutral densities and initial electron energy flux. Calculations have shown electron precipitation controls ionospheric plasma densities, with solar ionization in auroral regions playing only a secondary role due to the large zenith angles (Galand et al., 2011). Therefore, any realistic calculations of the high-latitude regions with GCMs require explicit inclusion of electron precipitation.

### 9.10. Concluding Remarks

One of the outstanding problems in the study of Saturn's atmosphere is the He/$H_2$ ratio as the mean molecular mass of the atmosphere is needed to correctly calculate the pressure levels as a function of radial distance. As noted the Cassini UVIS team has not reported to date any measurements of the He 58.4 nm line which could constrain this critical ratio. But this line is formed well above the He homopause shown in Figure 9. Thus one has to extrapolate the inferred ratio into the well mixed atmosphere with considerable uncertainty.

At the end of the Cassini Mission, known as the Grand Finale Tour, the last five orbits will be flybys through Saturn's thermosphere penetrating down to the sub nanobar level, but above the homopause and before a final fatal plunge into the atmosphere. The motivation is to take advantage of the onboard INMS that is capable of measuring the neutrals: $H_2$, He, for sure and possibly HD at these low pressures. With a spacecraft velocity ~ 30 km s$^{-1}$, the kinetic energy of $H_2$ molecules colliding with the spacecraft is ~ 8.8 eV, in excess of the 4.5 eV $H_2$ dissociation energy. If one of the $H_2$ proton nuclei is imparted more than 4.5 eV of kinetic energy in a collision with the instrument, the $H_2$ bond would be broken and $H_2$ could be undercounted relative to atomic He. The measurement of HD will be marginal if its actual mixing ratio were close to what is displayed in Figure 9.9.

But if one steps down sequentially in altitude during the last four orbits after spacecraft safety is confirmed from the first orbit, one can improve the chances of measuring HD and get better density profiles of $H_2$ and He. While performing the Grand Finale Tour, there will be many opportunities for UV stellar occultations to add more $H_2$ density profiles at a variety of latitudes to complement the more than 40 occultations discussed in Section 9.2.2. The challenge will be to translate He/$H_2$ and HD/$H_2$ density ratio profiles above the homopause into extrapolation of asymptotic values deep in the well-mixed atmosphere. For Jupiter, the latter was only achieved by dropping the Galileo Probe into the atmosphere and making measurements down to ~ 20 bar. In comparison to Jupiter our knowledge of the structure of Saturn's thermosphere will be far superior due to the large number solar and stellar occultations.

Far more certain, is that INMS will measure the composition of Saturn's ionosphere and, hopefully, obtain clear evidence of water group ions and infer their effect on electron densities. The ion composition will provide a consistency check on the neutral composition measurements.

One of the fundamental problems in understanding the thermospheres of Saturn and Jupiter is the heating mechanism(s) that accounts for their temperatures far exceeding what solar UV radiation can generate. Auroral heating is sufficient to solve this "energy crisis" for Saturn's thermospheric heating, if it can be efficiently redistributed to low latitudes. This solution has been rejected on the basis of thermospheric GCM calculations such as the STIM model. The net effect is considerably colder thermospheric temperatures than derived from occultations outside the polar regions. The fundamental cause of this under prediction by GCMs is that ion drag induces zonal winds in the retrograde, westward direction in the auroral regions, which when acted on by the Coriolis force generate poleward meridional winds that transport and confine auroral heating to polar latitudes rather than transport heat to the equator where it is most needed.

Almost all thermospheric GCMs addressing this problem include the assumption of hydrostatic equilibrium, which is only applicable to low Mach number flows of less than 0.3 (Kundu, 1990). But the STIM model produces neutral winds up to the sound speed in some regions and the hydrostatic assumption is no longer valid. The Coriolis force has a term in the radial direction of $2\Omega\ u\ \cos(lat)$, which cannot be ignored for flows in excess of Mach 0.5. Also, the hydrostatic approximation filters out acoustic-gravity waves which can transmit energy out of the polar regions to lower latitudes. Auroral heating is a spatial and time dependent forcing capable of generating such waves, which have horizontal group velocities close to the sound speed. Thus auroral power pulses can be propagated to the equator in less than two Saturn days. Likewise, supersonic ions **E x B** convecting through the auroral thermosphere must generate shocks in the neutral atmosphere, which cannot be handled by hydrostatic GCMs. Thus the energy "crisis" may not be inadequate total power input but a problem in the global redistribution of power in models. Contributing to this is a further shortcoming of all GCM simulations published to-date, the neglect of dynamical coupling to regions below in the form of mean background winds and upward propagating waves, an aspect that is known to considerably affect the circulation in the Earth's lower thermosphere. Work to address this shortcoming is underway with STIM and may lead to more realistic lower thermosphere dynamics which favor polar energy redistribution towards the equator.

Finally the thermal structure of the pressure region (~ 0.030-1 µbar) is not well characterized and the question of whether a mesopause is present is unanswered. A well instrumented probe(s) could address this question as well as shed light on the He/$H_2$ ratio and atmospheric structure from sub-nanobar to 10+ bars.

## Acknowledgments

DFS was supported by the Cassini-Huygens Mission through JPL Contract Nos. 1408487 and NASA Grant NNX10AB84G. TTK was supported by the NASA CDAPS Grant NNX14AD51G.

**Figures**

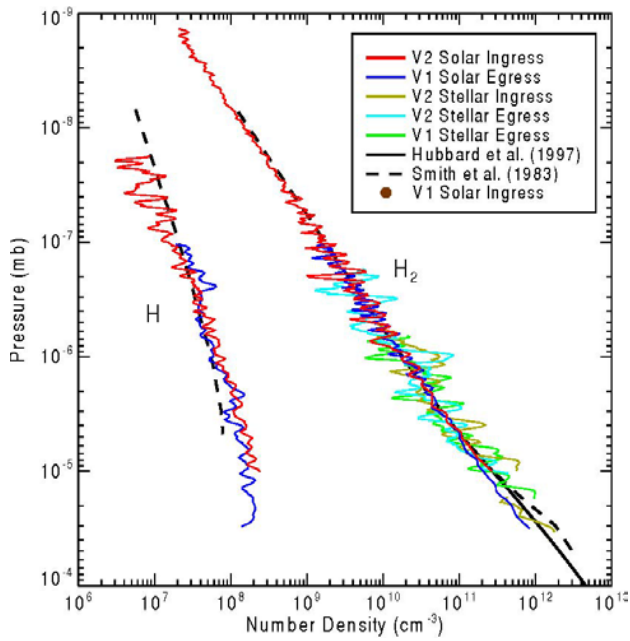

Figure 9.1. Density profiles of $H_2$ and H retrieved by Vervack and Moses (2015) from five low to mid latitude Voyager/UVS occultations.  The results are compared with previous retrieval by Smith et al. (1983) and the density profile retrieved by Hubbard et al. (1997) from a ground-based stellar occultation.  The figure is from Vervack and Moses (2015) and reprinted from Icarus with permission by Elsevier.

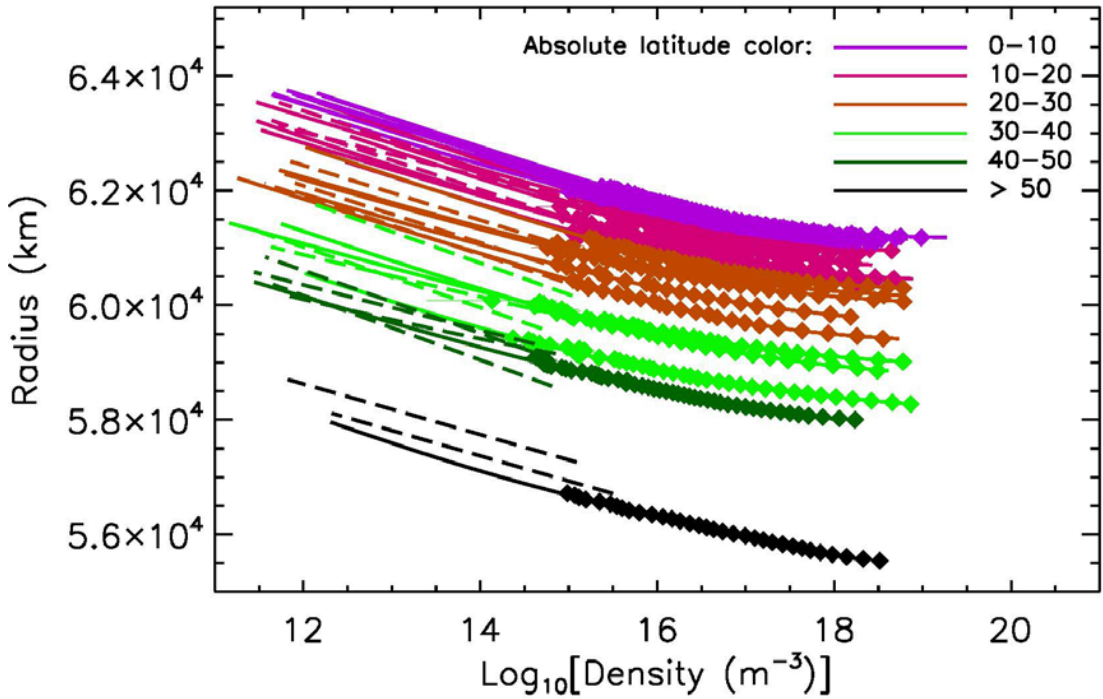

Figure 9.2. $H_2$ density as a function of radius (from the center of Saturn) based on the Cassini/UVIS stellar (solid lines and diamonds, Koskinen et al., 2015) and solar occultations (dashed lines, Koskinen et al., 2013). The color scale is based on latitude, which decreases from lower radial distances (black) to higher radial distances (purple). No distinction between northern and southern latitude is made. The data points (diamonds) show inverted densities while the solid and dashed lines show forward model density profiles. The figure is based on Koskinen et al. (2015).

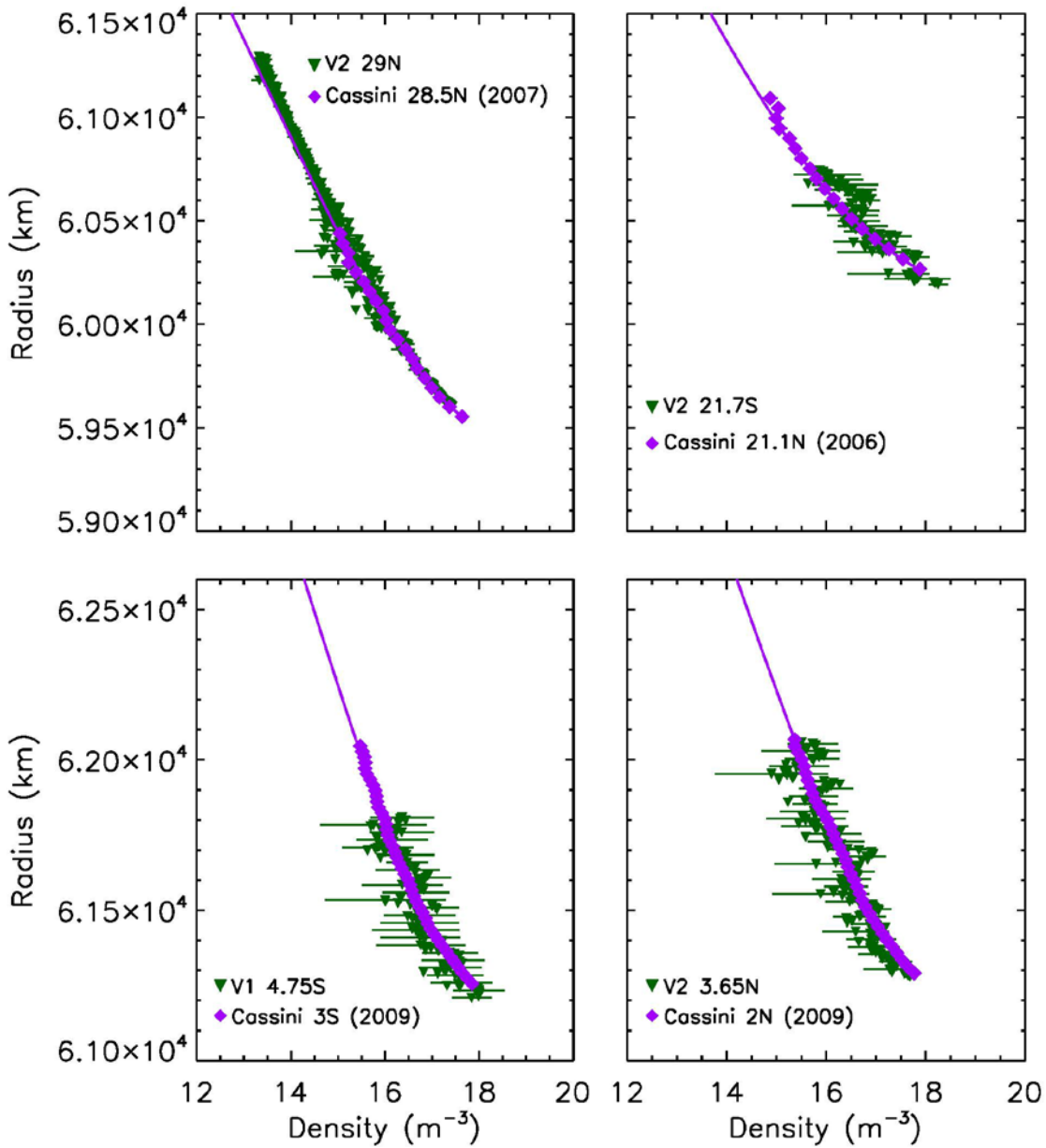

Figure 9.3. $H_2$ density profiles based on Cassini/UVIS stellar occultations (purple diamonds and solid lines, Koskinen et al., 2015) and Voyager/UVS solar (29N) and stellar occultations (green triangles, Vervack and Moses, 2015) at roughly equivalent planetocentric latitudes. The uncertainties in the Cassini profiles are mostly too small to be visible while the uncertainties in the Voyager profiles are larger. Note that the Cassini (2009) and Voyager V1 and V2 occultations were obtained during the equinox season.

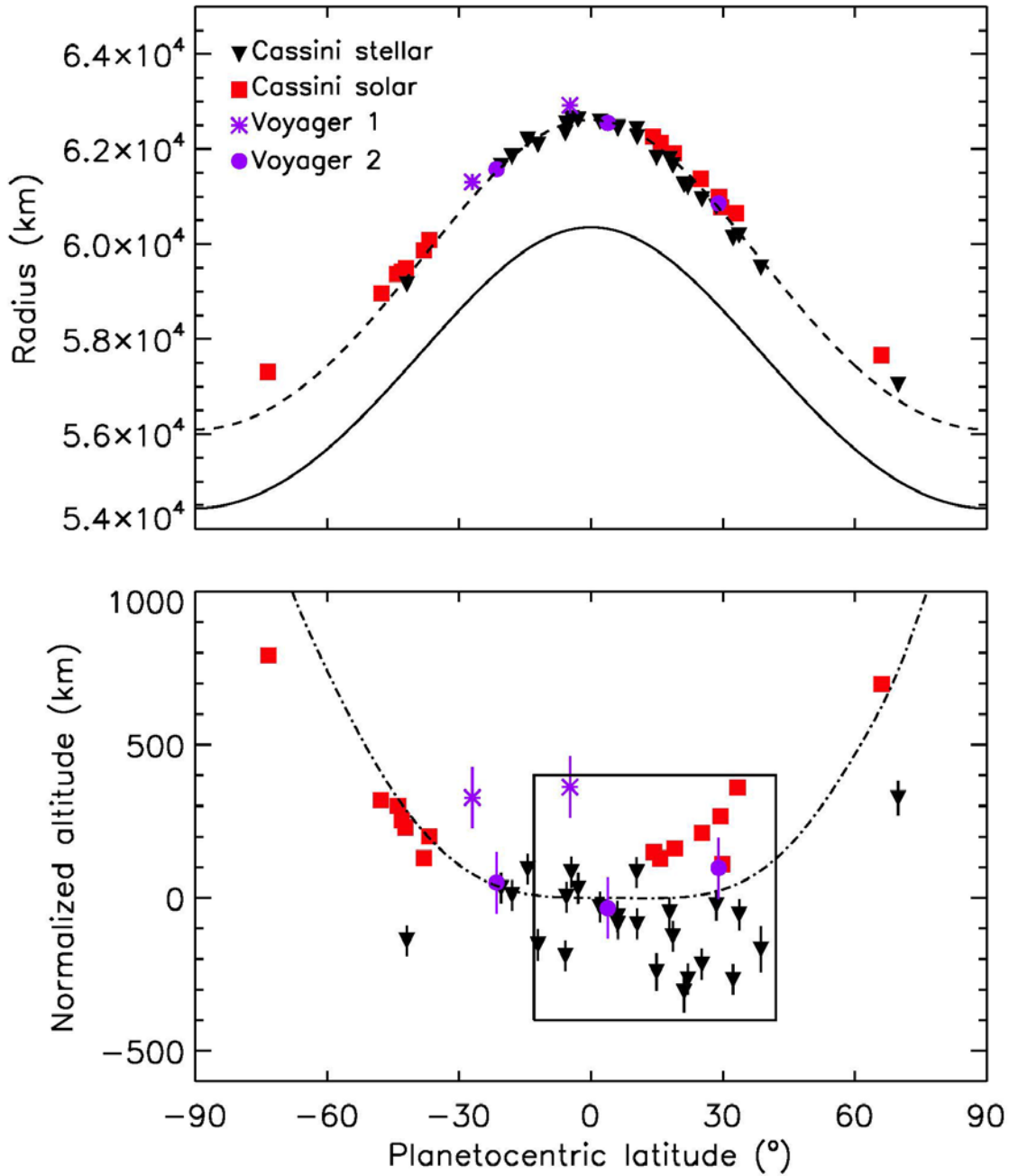

Figure 9.4. Upper panel: Radial distances to the 0.01 nbar pressure level based on Cassini/UVIS solar (red squares), Koskinen et al., 2013) and stellar (black diamonds, Koskinen et al., 2015) occultations as well as Voyager 1 and Voyager 2 occultations (green stars and filled circles, respectively, Vervack and Moses, 2015). The solid line is the 100 mbar reference level (Anderson and Schubert, 2007) and the dashed line is an extrapolation of this reference model to 0.01 nbar that matches the equatorial occultations

from 2008/2009. Lower panel: altitude of the data points relative to the 0.01 nbar reference level (dashed line in the upper panel). The square indicates data points included in Figure 9.5. The dashed-dotted line shows normalized altitudes predicted by the new results from the GCM of Müller-Wodarg et al. (2012) (see Section 9.8.3). The figure was taken from Koskinen et al. (2015).

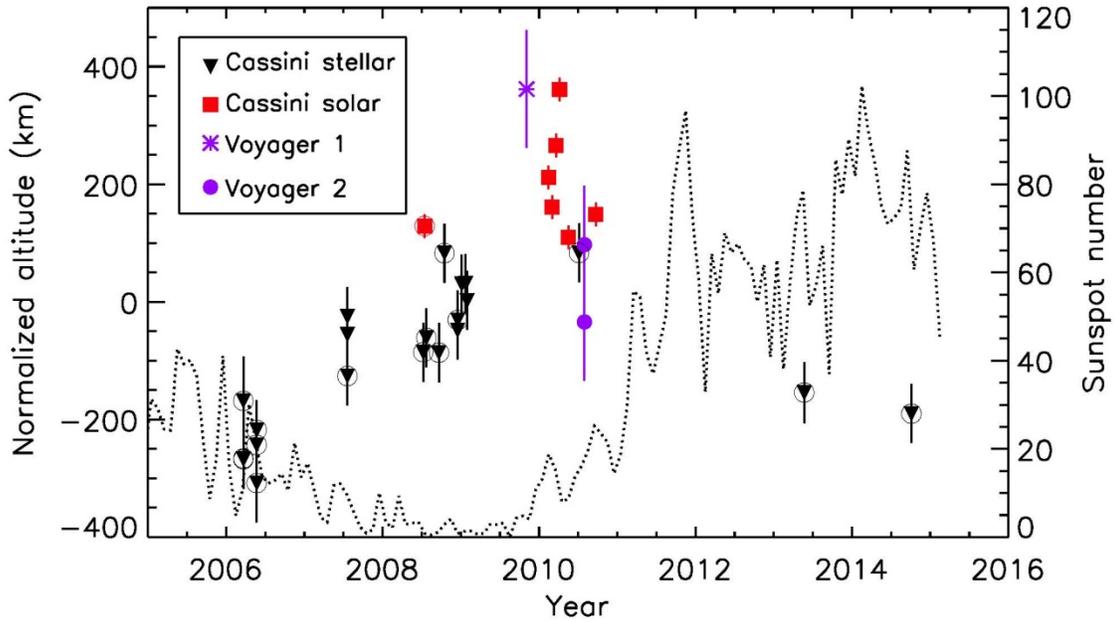

Figure 9.5. Normalized altitude above and below the 0.01 nbar reference model as a function of time, including the data points inside the square in the lower panel of Figure 9.4. The Voyager/UVS data points (based on Vervack and Moses, 2015) are shown at the equivalent season and time after the equinox. The Cassini occultations that fall either into the ring shadow or probe the latitudes of the ring shadow are indicated by open circles. The sunspot number is shown by the dotted line. The figure is based on Koskinen et al. (2015).

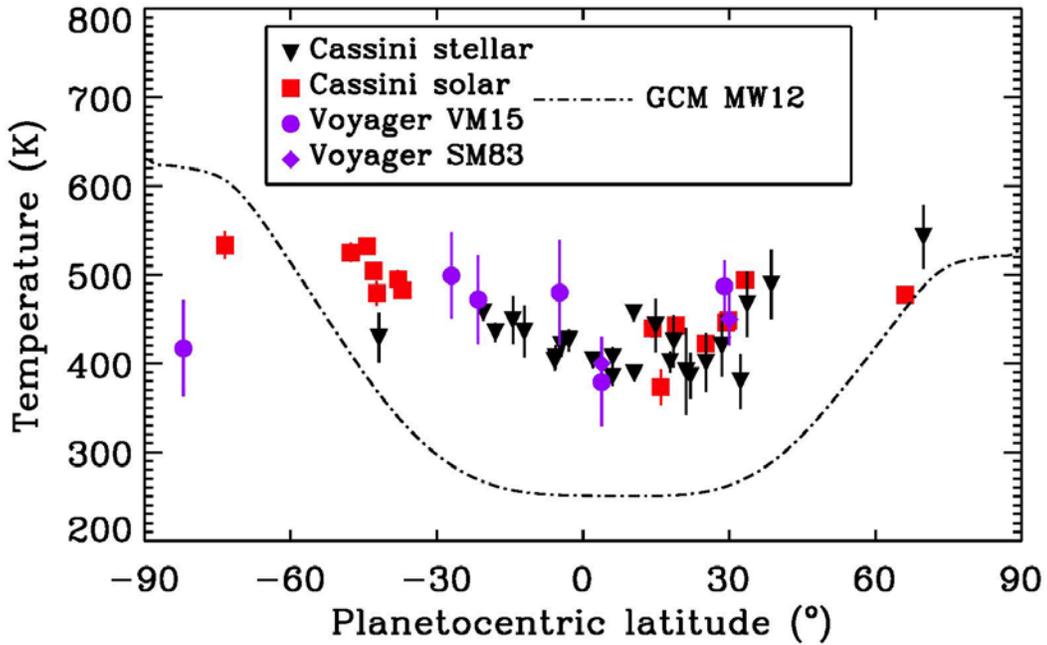

Figure 9.6. Exospheric temperatures retrieved from Cassini/UVIS solar (purple diamonds, Koskinen et al., 2013) and stellar (black triangles, Koskinen et al., 2015) occultations, and Voyager/UVS occultations by Smith et al. (1983) (green squares) and Vervack and Moses (2015) (green circles). The dashed line shows the exospheric temperatures based on new results from the GCM of Müller-Wodarg et al. (2012) (see Section 9.8.3). The figure was taken from Koskinen et al. (2015).

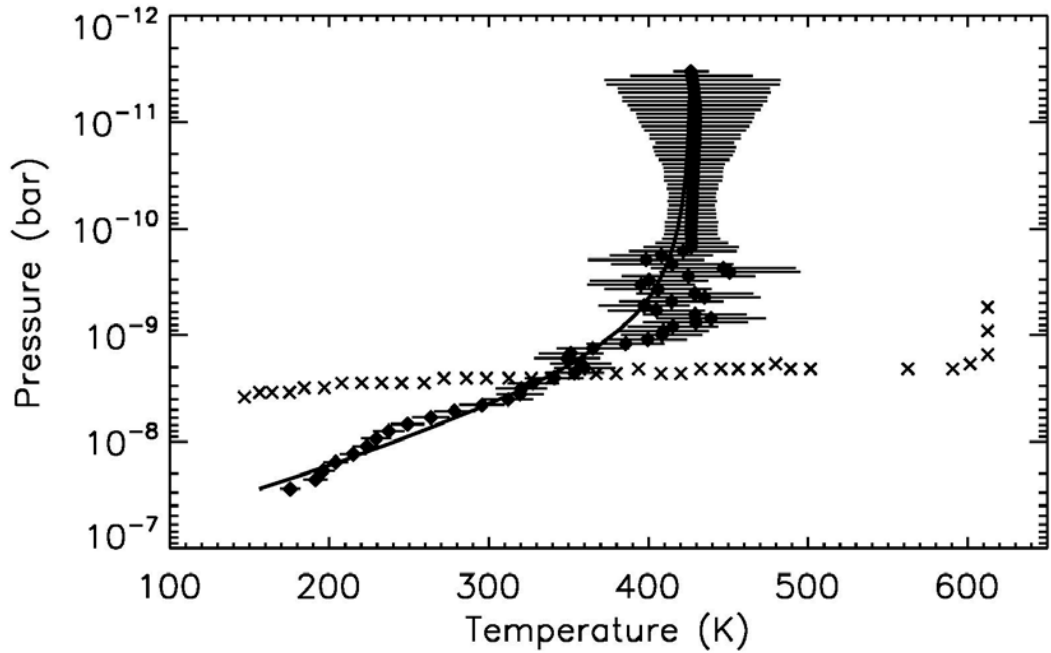

Figure 9.7. Temperature-pressure (T-P) profile based on the occultation of β Crucis from January 2009 that probes the atmosphere at the planetocentric latitude of 3°S. The diamonds and solid line show the direct retrieval and forward model profiles from Koskinen et al.(2015) while the crosses show the temperature profile from Shemansky and Liu (2012) (see text).

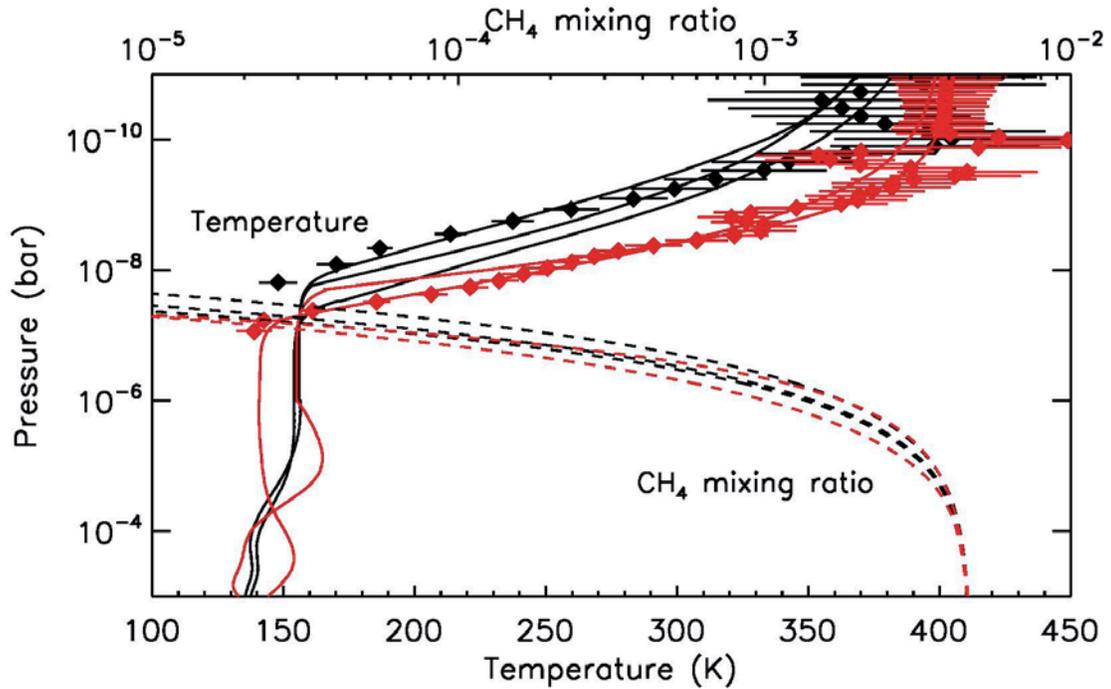

Figure 9.8. Forward model temperature profiles (solid lines) and CH₄ mixing ratio profiles (dashed lines) for spring of 2006 (black) and December 2008 (red). The occultations probe the atmosphere at planetographic latitudes of 2 – 20°N (see text). Diamonds show the direct retrieval temperatures for two of the occultations. The figure was taken from Koskinen et al. (2015).

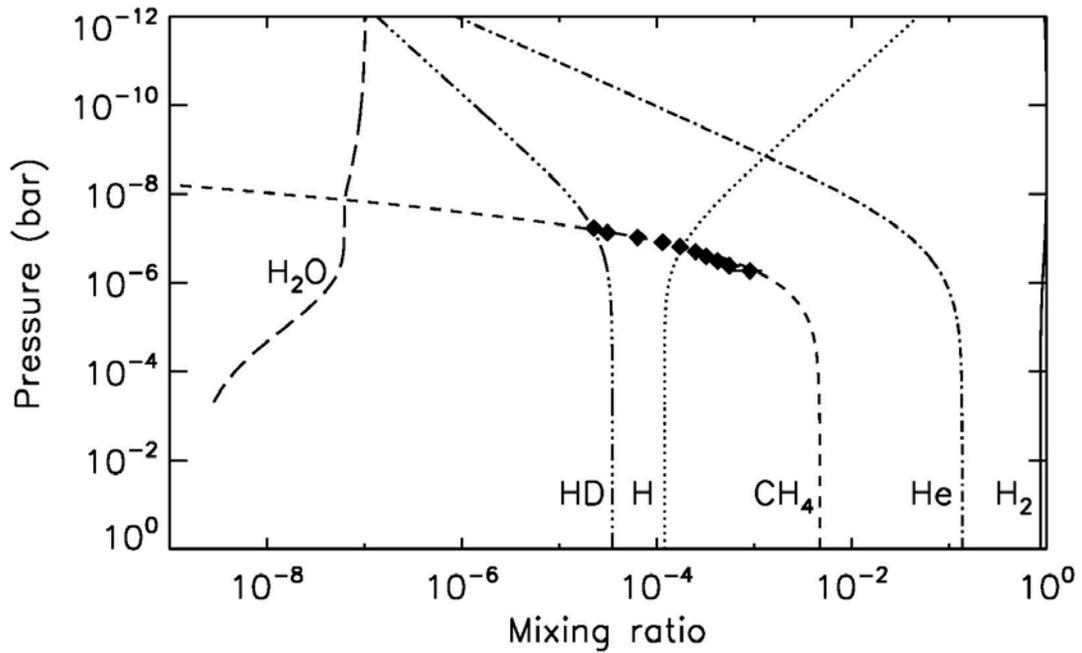

Figure 9.9. Calculated volume mixing ratio profiles that correspond to the atmospheric structure model (temperature-pressure profile) based on the occultation of ε Eridani near the planetographic latitude of 25°N from spring of 2006 and nearly coinciding CIRS limb observations (see Figure 9.8). Diamonds show the $CH_4$ mixing ratios retrieved from the occultation to constrain the $K_{zz}$ profile. The assumed mole fractions at the 1 bar level are 0.1355 for He, 4.7 x $10^{-3}$ for $CH_4$ and 3.5 x $10^{-5}$ for HD. For water we assumed an influx of $10^6$ $cm^{-2}$ $s^{-1}$ and fixed the mixing ratio to 3 x $10^{-9}$ at 0.5 mbar based on recent Herschel observations (Fletcher et al., 2012). The mixing ratio of H was set to match the upper limit of 5 % in the thermosphere (Koskinen et al., 2013).

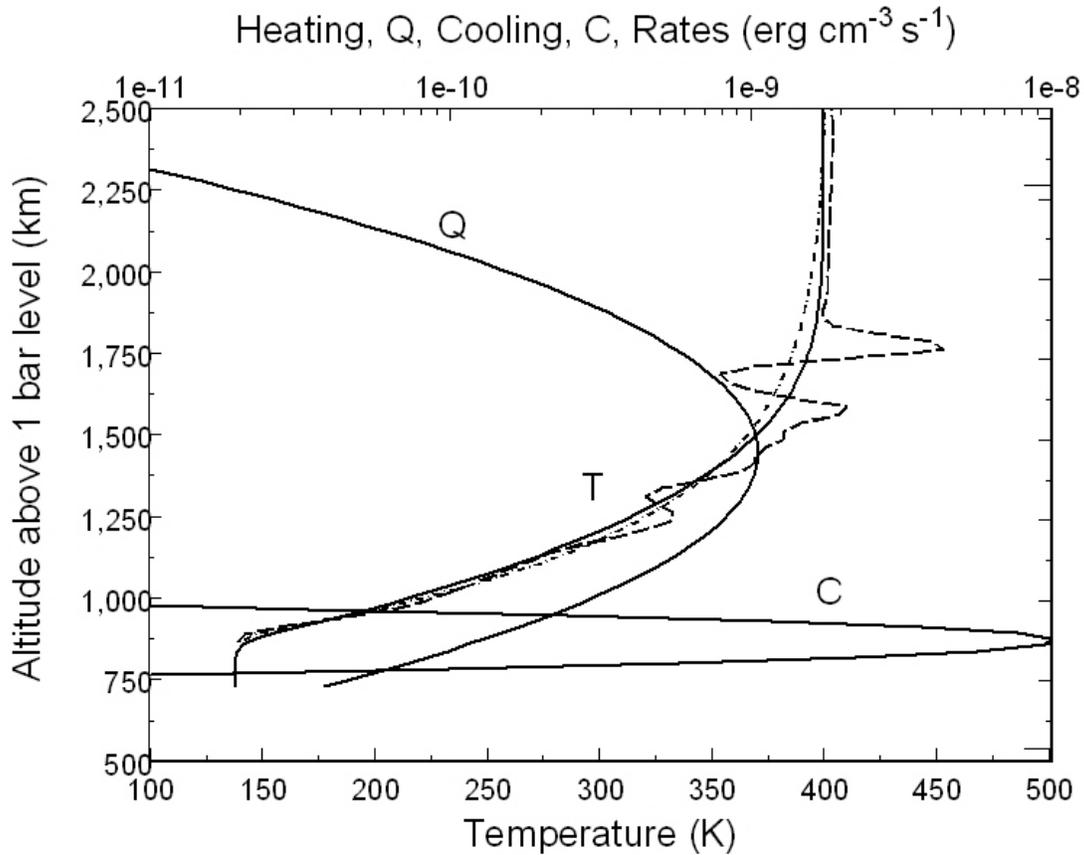

Figure 9.10. Inferred heating, Q, and cooling, C, rates from December 2008 stellar occultation derived temperature profiles illustrated in Figure 9.8 (red diamonds, there; here dashed line) and from forward model (solid red lines, there; here dash dot line). Solving the heat conduction equation with the inferred Q and C profiles yields the solid line temperature. The inferred net integrated heating rate is 0.072 ergs cm$^{-2}$ s$^{-1}$, with peak heating at 1450 km and 0.65 nbar, while the peak cooling is inferred at 870 km and 70 nbar.

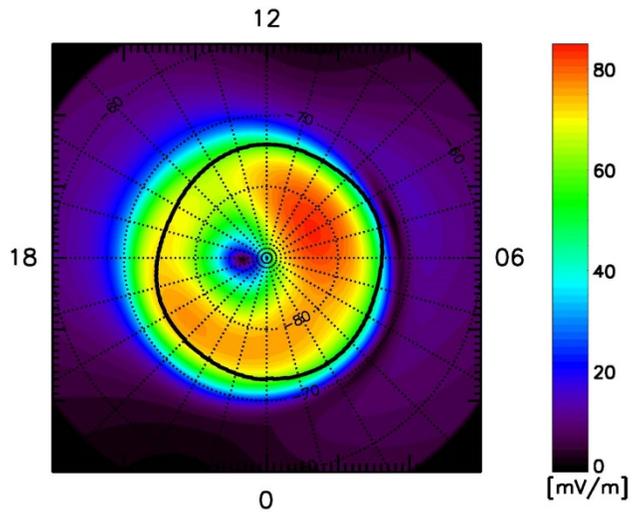

Figure 9.11. The magnetospheric electric field strength (color contours) as mapped into the southern polar region with the locations of maximum field-aligned current (black symbols), which coincide with the regions of largest electron precipitation into Saturn's polar upper atmosphere. The figure axes indicate local times.

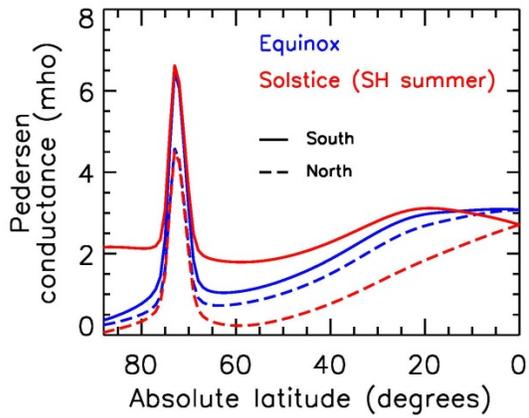

Figure 9.12. Zonally averaged Pedersen conductances in Saturn's ionosphere as a function of latitude. Solid lines denote the southern hemisphere values, dashed values are for the northern hemisphere. The blue lines are for an equinox simulation of STIM and the red lines for southern hemisphere summer conditions.

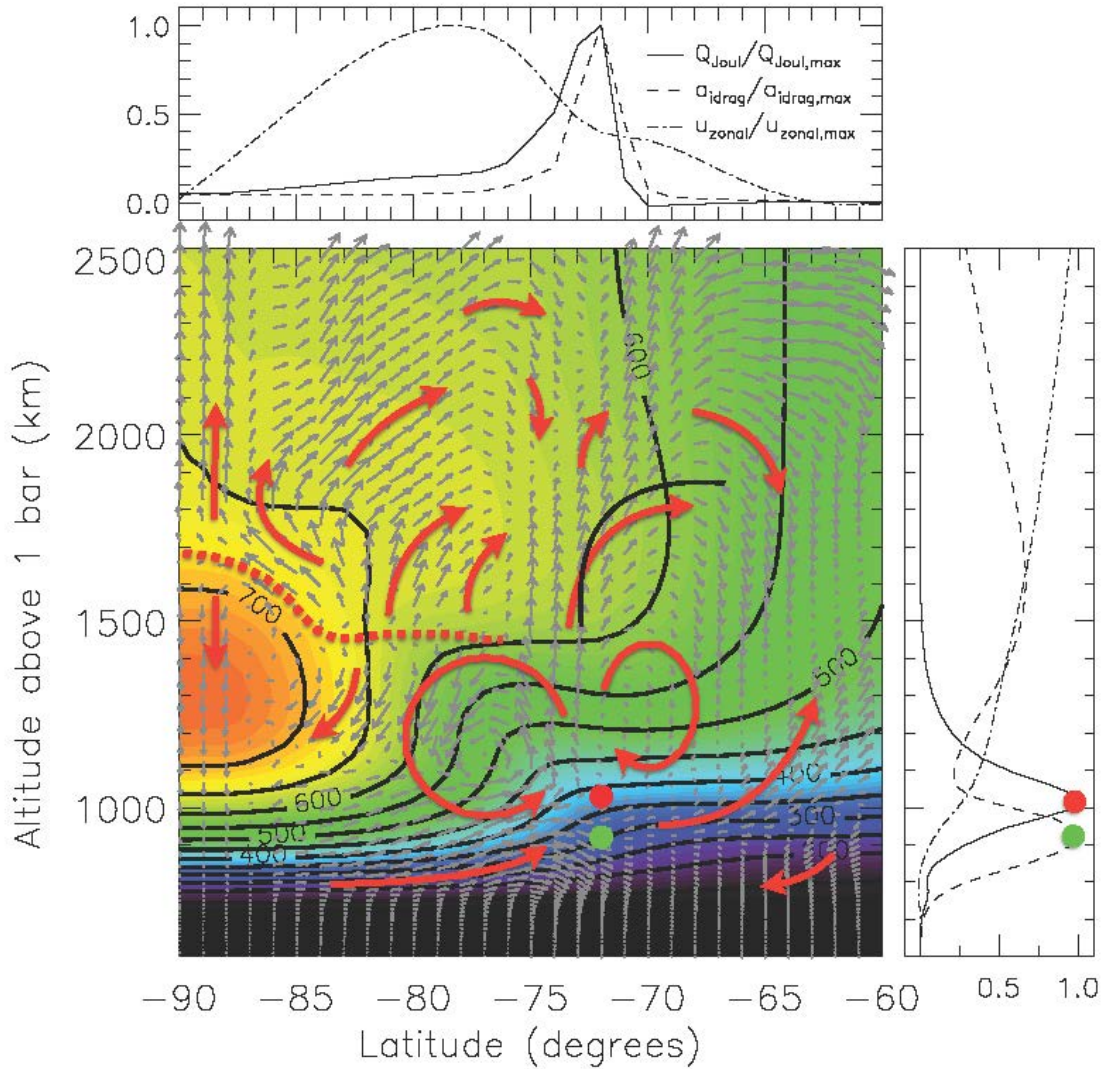

Figure 9.13. Neutral temperatures contours and meridional/vertical circulation wind vectors (grey arrows) as simulated by STIM for equinoctial conditions. The longest grey arrow corresponds to around 350 m s$^{-1}$. Red arrows illustrate the broad circulation pattern and the red dotted line separates regions of upward and downward vertical winds. Two line plot panels with normalized quantities show Joule heating rates (solid line, $Q_{Joule}/Q_{Joule,max}$), ion drag acceleration (dashed, $a_{idrag}/a_{idrag,max}$) and zonal wind velocities (dashed-dotted, $u_{zonal}/u_{zonal,max}$). The right panel shows vertical profiles at 72°S latitude while the curves in the top panel are latitudinal profiles at the height of peak Joule heating (solid line) and at the height of peak ion drag (dashed and dashed-dotted lines). The red dot in the temperature panel denotes the location of maximum Joule heating shown in the line plot, also labeled there with a red dot. The green dots on the temperature panel and line plot denote the region of maximum zonal ion drag. Peak Joule heating and ion drag both occur at 72°S latitude but shifted vertically by ~100 km from one another. Zonal winds in the upper panel are normalized to a value of 334 m s$^{-1}$; their largest values in the right panel reach 1500 m s$^{-1}$. The peak values of zonal ion drag and Joule heating in the two panels are 0.02 m s$^{-2}$ and 1.9x10$^{-8}$ W m$^{-3}$, respectively.